\documentclass[aps,prd,nofootinbib,twocolumn,notitlepage]{revtex4-2}
\pdfoutput=1
\usepackage[T1]{fontenc}
\usepackage{amsmath,amssymb}
\usepackage{graphicx}
\usepackage{bm}
\usepackage{xcolor}
\usepackage[colorlinks=true, linkcolor=blue, citecolor=blue, urlcolor=blue]{hyperref}
\usepackage{orcidlink}
\usepackage{wrapfig}
\usepackage[normalem]{ulem}
\definecolor{light-gray}{gray}{0.95}
\usepackage{tcolorbox}
\usepackage{float}
\usepackage{enumitem}
\usepackage{subcaption}
\usepackage{caption}
\def\be{\begin{equation}}
	\def\ee{\end{equation}}
\def\beq{\begin{eqnarray}}
	\def\eeq{\end{eqnarray}}
\def\bmig{\begin{figure}}
	\def\efig{\end{figure}}
\def\lt{\left}
\def\rt{\right}
\newcommand{\nn}{\nonumber}
\newcommand{\bib}{\begin{thebibliography}}
	\newcommand{\ebib}{\end{thebibliography}}
\renewcommand{\nn}{\nonumber}

\begin{document}
	
	\title{Resonant Dynamics of Gravitational Wave and Chiral Alfv\'en wave in the Early Universe}
	
    \author{$^1$\href{https://arunp77.github.io/}{Arun Kumar Pandey}\,
    \orcidlink{0000-0002-1334-043X}}
    
    \author{$^2$Subalakshmi A\, \orcidlink{https://orcid.org/0009-0003-9416-4267}}
    \email{subalakshmiarun00@gmail.com}
    
    \author{$^2$ Sampurn Anand\,\orcidlink{0000-0003-4346-6276}}\thanks{
    \href{mailto:sampurn@acad.cutn.ac.in}{sampurn@acad.cutn.ac.in}}
    
    \affiliation{$^1$Department of Physics and Astrophysics, University of Delhi, Delhi 110 007, India}
    \affiliation{$^2$Department of Physics, Central University of Tamil Nadu, Thiruvarur 610 005, Tamil Nadu, India}
	
	\begin{abstract}
		In this work, we investigate the resonant interaction between stochastic gravitational waves (GWs) and chiral Alfv\'en waves in a magnetized chiral plasma in the early Universe. Starting from covariant chiral magnetohydrodynamics coupled to linearized gravity, we derive a closed system of equations for the coupled chiral-Alfv\'en velocity and magnetic-field perturbations. Both analytics and numerics show a parametric resonance at the sum frequency of the two CVE-split branches, with an instability band that widens as the GW strain grows. This CVE-to-Alfv\'en ratio is not a free parameter. In the early Universe, it is fixed by the Standard Model's relativistic degrees of freedom. Including the plasma's backreaction on the GW makes the energy exchange nonlinear, breaking the usual single-frequency Manley-Rowe picture. For a self-consistent choice of parameters, where backreaction is only a small correction, the unsuppressed resonance still drives the system to a finite-time blow-up within a few Hubble times. We then work out the resonant frequencies and growth rates involved, mapping out what current and future gravitational-wave detectors could observe.
	\end{abstract}
    
	
	\maketitle
	\section{Introduction}\label{sec:intro}
	Chirality is a fundamental attribute of relativistic fermions, and its macroscopic consequences are of interest across nuclear, particle, condensed matter systems, and astroparticle physics. Within the Standard Model, the Adler-Bell-Jackiw anomaly \cite{adler1969axial, bell1969pcac} and the associated 't~Hooft vertex \cite{t1976symmetry} allow chirality-violating processes to proceed even though the underlying gauge interactions separately conserve the current in each chirality sector. In the early Universe, such processes can generate a chiral asymmetry, an excess of right- over left-handed fermions quantified by the chiral chemical potential $\mu_5=(\mu_R-\mu_L)/2$, either as a by-product of baryogenesis or through an independent mechanism. Coupling chiral asymmetry to a hot, magnetized plasma generates two anomalous transport phenomena that lack parity-invariant analogs. First, the chiral magnetic effect (CME), which induces an electric current parallel to an applied magnetic field \cite{vilenkin1980equilibrium,fukushima2008chiral}, and second, the chiral vortical effect (CVE), which drives a current along the fluid vorticity \cite{son2009hydrodynamics,landsteiner2011gravitational} (see Ref.~\cite{kharzeev2016chiral}). When the electromagnetic field is itself dynamical, a chiral plasma with $\mu_5\neq0$ develops an instability called the chiral plasma instability (CPI), in which helical magnetic modes grow exponentially at the expense of the chiral charge \cite{joyce1997primordial, akamatsu2013chiral, boyarsky2012self}. The same anomalous currents also reshape the plasma's own linear wave spectrum in the presence of a background field, giving rise to normal modes distinct from ordinary MHD, including in dissipative extensions \cite{pandey2018effectbackgroundmagneticfield}. These anomalous transport effects have since been invoked to address several open problems in early-Universe cosmology, most notably the origin of the observed and inferred cosmological magnetic fields \cite{tashiro2012chiral,bhatt2016primordial,pandey2015primordial,durrer2013cosmological,anand2017chiral} (see Ref.~\cite{subramanian2016origin, kamada2023chiral} for a recent review of chiral effects in astrophysics and cosmology).
	
	On the other hand, the primordial stochastic gravitational-wave (GW) background is a generic prediction of inflationary cosmology \cite{grishchuk1975amplification,starobinsky1979spectrum,abbott1984constraints}. In vacuum, a linearized GW does not couple to an electromagnetic or magnetohydrodynamic (MHD) wave propagating in the same direction. A background magnetic field is required to lift this decoupling and allow the two to exchange energy resonantly. This GW-magnetoplasma coupling has been studied extensively for ordinary (non-chiral) plasmas 
    \cite {marklund1999interaction,servin2000parametric,servin2003resonant,kallberg2004nonlinear,brodin2001photon}. Also,  chiral MHD turbulence has been shown to source a circularly-polarized GW background \cite{brandenburg2021relic,brandenburg2024relic,solodukhin2024beltrami}. The resonant response of a chiral plasma to an externally imposed GW has not been investigated in a plasma that supports the anomalous CME/CVE currents described above. Since the early Universe is plausibly both chirally asymmetric and magnetized, and a stochastic GW background is expected to have been present throughout, it is natural to ask whether the resonant coupling between a GW and an MHD wave persists and, if so, how it is modified once the plasma's chiral transport is retained. We have addressed precisely that question in this work.
	
	The remainder of the paper is organized as follows: Section~\ref{sec:-II} describes the linearized metric tensor perturbation and introduces the tetrad (orthonormal-frame) formalism used throughout. It also sets out the covariant equations of motion for Chiral MHD (ChMHD) coupled to a weak GW and derives the frame components of the generalized Maxwell equations and the GW-induced force on the fluid. Section~\ref{sec-GW-wave-inter} is dedicated to a GW propagating along the background magnetic field, derives the master equations coupling the chiral-Alfv\'en polarizations to the magnetic-perturbation polarizations, and derives the leading backreaction of the plasma on the GW. Section~\ref{sec-results} presents the findings of numerical analysis of the master equations, the resulting parametric-resonance instability, its dependence on the plasma's chirality, and on the production epoch. This section also addresses the question of backreaction dynamics, and a physically anchored early-Universe physically motivated parameter set. We conclude with a summary of our main findings in Section~\ref{sec-conclusion}.
    \section{Basic equations governing the dynamics of the  chiral plasma and GW}
    \label{sec:-II}
    \subsection{Gravitational wave evolution}

    To describe gravitational waves, we perturb the metric about a flat background, given by
    \be
        g_{ab} = \eta_{ab} + h_{ab}, \qquad |h_{ab}| \ll 1\, ,
        \label{eq:metric_ptbn}
    \ee
    where $\eta_{ab}$ is the Minkowski metric and $h_{ab}$ is the perturbation. The numerical values of tensor components depend on the choice of reference frame, so fixing a frame breaks manifest covariance under coordinate transformations. This is nevertheless the standard way to remove the spurious gauge degrees of freedom and expose the physical content of the theory. Once a frame in which Eq.~\eqref{eq:metric_ptbn} holds has been chosen, a residual gauge freedom remains. Working in the gauge $\partial^b \bar h_{ab} = 0$, with $\bar h_{ab} = h_{ab} - \tfrac{1}{2}\eta_{ab}h$, the wave equation takes the form
	\be
    	\Box \bar h_{ab} =-16 \pi G\, \delta {\cal T}_{ab} \, ,
    	\label{eq:gw-prop}
	\ee
	where $\delta{\cal T}_{ab}$ is the source term and $\Box$ is the d'Alembertian. To eliminate the remaining gauge freedom, we work in the transverse-traceless (TT) gauge,
	\be
	   h^{0a} = 0, ~~~~ h^i{}_i = 0, ~~~~ \partial^j h_{ij} = 0\, .
	\ee
	In this gauge, the metric of a linearized gravitational wave propagating along $z$ is
	\begin{align}
		ds^2 = -dt^2 & ~+~  (1+h_+)~dx^2 ~+~   (1-h_+) ~dy^2 \nonumber \\
		& + 2~h_\times ~dx~dy~ +~dz^2
		\label{eq:metric-pertub}
	\end{align}
	where $h_+$ and $h_\times$ are the two polarization amplitudes, $|h_+|,|h_\times|\ll1$. In vacuum, $h_+$ and $h_\times$ depend on the single retarded-time variable $z-t$. The weak interaction with the plasma considered here perturbs this only at higher order, so we retain $h_{+,\times}=h_{+,\times}(z-t)$ and hence $\partial_z=-\partial_t$ is used throughout \cite{servin2003resonant, mofiz2007generation}.
    
	To proceed further, it is convenient to introduce an orthonormal (tetrad) frame $\{{\bf e}_a\}_{a=0,1,2,3}$ related to the coordinate vector fields $\partial_\mu$ by \cite{kallberg2004nonlinear,brodin2001photon,ellis2002cosmological},
	\[ 
	{\bf e}_a = e_a^\mu \partial_\mu \quad \Leftrightarrow\quad {\bf e}_a (f) = e_a^\mu \partial_\mu(f)\, .
	\] 
	This is simply a change of basis and induces the corresponding change of tensor components. To linear order in $h$, the contravariant tetrad corresponding to the metric~\eqref{eq:metric-pertub} is
	\begin{align}
		{\bf e}_0 &= \partial_t,  \quad {\bf e}_1 = \left(1 -\frac{h_+}{2}\right)\partial_x - \frac{h_\times}{2} \partial_y \nn \\
		{\bf e}_2 &= - \frac{h_\times}{2} \partial_x + \left(1 +\frac{h_+}{2}\right)\partial_y,  \quad {\bf e}_3 = \partial_z . 
		\label{eq:tetrad}
	\end{align}
    Substituting the metric~\eqref{eq:metric-pertub} and tetrad~\eqref{eq:tetrad} into the linearized Einstein equation~\eqref{eq:gw-prop}, then subtracting the former and adding the latter, results in
	\begin{eqnarray}
		\Box  h_+ = -8 \pi G\, (\delta{\cal T}_{11}- \delta{\cal T}_{22})\, ,\label{eq:lEeq1} \\
		\Box h_\times = -8 \pi G\, (\delta{\cal T}_{12}+ \delta{\cal T}_{21})~\label{eq:lEeq2}\, .
	\end{eqnarray}
	In the absence of a source the right-hand side vanishes and the GW propagates as a plane wave while a nonzero $\delta{\cal T}_{11}-\delta{\cal T}_{22}$ or $\delta{\cal T}_{12}+\delta{\cal T}_{21}$ drives or damps the corresponding polarization. Throughout this work, we retain only terms linear in $h$.
	\subsection{Chiral plasma}
	The dynamical evolution of the plasma is described within relativistic hydrodynamics. In a chiral plasma at finite chemical potential $\mu=(\mu_R+\mu_L)/2$ and chiral chemical potential $\mu_5=(\mu_R-\mu_L)/2$, the total electric current $j^a=j^a_L+j^a_R$ is conserved, while the axial current $j^{a5}=j^a_R-j^a_L$ is anomalous. The chiral hydrodynamic equations, together with energy-momentum conservation, read
	\begin{align}
		\nabla_a~T^{ab} &= 0  \label{eq:conserv-2}\\
		\nabla_a j^a & = 0 \\
		\nabla_a j^{a5} &= -C\,{\bf E}\cdot {\bf B} \label{eq:conserv-1}
	\end{align}
	Here $j^a$ is the (vector) electric current, $j^{a5}$ is the axial (chiral) current, $E_a=F_{ab}u^b$, $B_a=\tfrac12\epsilon_{abc}F^{bc}$, and $u^a=(\gamma,\gamma{\bf v})$, with $\bf v$ the local fluid three-velocity and $\gamma=(1-v^2)^{-1/2}$ the associated Lorentz factor; $F^{ab}$ is the electromagnetic field-strength tensor and $\nabla$ denotes covariant differentiation. We use the mostly-plus signature $(-,+,+,+)$ throughout.
	
	The energy-momentum tensor receives contributions from the fluid and the electromagnetic field, $T_{ab}=T_{ab}^{\rm f}+T_{ab}^{\rm em}$, with
	\begin{align}
		T_{ab}^{\rm f} = (\epsilon + p) u_a\,u_b + p\,g_{ab} + \pi_{ab} \label{eq:StressCons}
	\end{align}
	where $\epsilon$, $p$, and $u^a$ are the energy density, pressure, and four-velocity, respectively, $g_{ab}$ is the metric tensor, and the dissipative stress $\pi_{ab}$ is neglected throughout. The electromagnetic part is
	\begin{align*}
		T_{ab}^{\rm em} =  F_{a}^c F_{bc}-\frac{1}{4} g_{ab}\, F^{cd}F_{cd}\, .
	\end{align*}
	The vector and axial current densities are
	\begin{align*}
		j^a  &=  n\,u^a + \xi_B\, B^a\,  + \xi \omega^a , \\
		j^{a5} &=  n_5\,u^a + \xi^5_B\, B^a\,  + \xi^5 \omega^a
	\end{align*}
	where $n$ is the charge density and $\omega^a=\epsilon^{abcd}u_b\partial_cu_d$ is the vorticity. The transport coefficients $\xi_B$ and $\xi$ correspond, respectively, to the chiral magnetic effect (CME) and the chiral vortical effect (CVE). The form of the transport coefficients is constrained by the second law of thermodynamics, $\partial_\mu s^\mu \geq 0$, in the presence of chiral and gauge-gravitational anomalies. Following the derivation in Refs.~\cite{son2009hydrodynamics,landsteiner2011gravitational,neiman2011relativistic}, they are given by
	\begin{eqnarray}
		\xi & = & C\mu^2\lt[1-\frac{2n\mu}{3(\epsilon + p)}\rt] +
		\frac{DT^2}{2}\lt[1-\frac{2n\mu}{(\epsilon + p)}\rt]
		\label{eq:xi} \\
		\xi_B & = & C\mu\lt[1-\frac{n\mu}{2(\epsilon + p)}\rt] -
		\frac{D}{2}\lt[\frac{nT^2}{(\epsilon + p)}\rt],
		\label{eq:xiB}
	\end{eqnarray}
	where $C=1/4\pi^2$ and $D=1/12$ denote the chiral anomaly and gauge-gravitational anomaly coefficients, respectively. In the high-temperature regime of the early Universe, $\mu/T \ll 1$, the coefficient $\xi$ approaches $(D/2)T^2$, becoming independent of the chiral imbalance. This term alone can source a primordial magnetic field even in the absence of a chiral charge~\cite{anand2017chiral}.
	
	It is important to note that Eqs.~\eqref{eq:xi} and~\eqref{eq:xiB} are adopted from a single-fermion-species framework ~\cite{son2009hydrodynamics,landsteiner2011gravitational,neiman2011relativistic,yamamoto2015chiral}.     In that setup, only one conserved charge exists, and thus only one chemical potential is defined. Consequently, there is no distinction between an ordinary chemical potential $\mu$ and a chiral chemical potential $\mu_5$, since a single fermion population, e.g., purely right-handed, does not allow for such a separation. The $(D/2)T^2$ term used throughout this work as $v_T \equiv \xi_0 v_A / \sqrt{\epsilon_0 + p_0}$ corresponds exactly to the quantity validated by Yamamoto in the context of chiral Alfv\'en waves~\cite{yamamoto2015chiral}. This contribution is robust, as it persists at $\mu=0$ for a genuinely chiral single-handed population, driven solely by the gravitational anomaly. However, $\xi_B$ and hence the chiral magnetic effect (CME) are negligible in the more realistic early-Universe scenario of comparable left- and right-handed populations with a small asymmetry $\mu_5 \ll T$ and need detailed investigation. The applicability of Eqs.~\eqref{eq:xi} and~\eqref{eq:xiB} to a near-symmetric two-species system is not guaranteed a priori. Determining whether $\xi_B$ genuinely decouples in this regime, or remains an independent channel alongside $\xi$, requires a multi-species framework that incorporates independent chemical potentials, $\mu$ and $\mu_5$. Such a framework could utilize the Kubo-formula approach of Ref.~\cite{landsteiner2011gravitational} or a chiral magnetohydrodynamic (ChMHD) induction equation driven linearly by $\Delta\mu = \mu_L - \mu_R$, independent of an ordinary chemical potential~\cite{boyarsky2012self}. We leave this investigation for future work. Importantly, this open question does not affect the validated chiral vortical effect (CVE) physics governed by $\xi$, which underpins the remainder of this study.
	
	Throughout this work, $\mu$ and $T$ (hence $\xi$ and $\xi_B$) are treated as fixed, homogeneous background quantities, while only the velocity, electromagnetic fields, and gravitational-wave-sourced metric perturbations are evolved dynamically. This treatment is justified because $\mu$ itself evolves only through the chiral anomaly equation~\eqref{eq:conserv-1}, sourced by $\mathbf{E}\cdot\mathbf{B}$ and diluted by Hubble expansion, both acting on the cosmological timescale $H^{-1}$. The resonance mechanism studied in this paper is self-consistent once many oscillation periods have elapsed per Hubble time, i.e., $ f_A, f_g \gg H$. Over this rapid timescale, keeping $\mu$ frozen is simply the leading-order equivalent of treating $n_0$, $\epsilon_0$, $p_0$, and $\mathbf{B}_0$ as uniform background fields. One consequence of this approximation is that the resonantly amplified velocity and field perturbations studied here do not feed back on $\mu$ through $\mathbf{E}\cdot\mathbf{B} \neq 0$. 
    
	The covariant Maxwell equations are defined as
	\begin{eqnarray}
		\nabla_a F^{ab} & = & j^b \label{eq:maxwell-source},\\
		\nabla_a F_{bc } + \nabla_b F_{ca } +  \nabla_c F_{ab }& = & 0 \label{eq:Bianchi-identity1}.
	\end{eqnarray}
	Following Refs.~\cite{kallberg2004nonlinear,brodin2001photon,ellis2002cosmological}, we introduce an orthonormal frame with basis vectors ${\bf e}_a=e^\mu_a\partial_\mu$ and write $\bar\nabla=({\bf e}_1,{\bf e}_2,{\bf e}_3)$ for the spatial gradient operator in this frame. Maxwell's equations in the $\{{\bf e}_a\}$ basis are given by
	\begin{align}
		\bar \nabla \cdot {\bf E} &= \rho + \rho_{_E} \label{eq:mx1}\\
		\bar\nabla\cdot{\bf B} & = \rho_{_B} \label{eq:mx2} \\
		{\bf e}_0 {\bf E} - \bar \nabla \times {\bf B} &= -{\bf j} - {\bf j}_{_E} \label{eq:max3} \\
		{\bf e}_0 {\bf B} + \bar \nabla \times {\bf E} &= -{\bf j}_{_B} \label{eq:max4}\, .
	\end{align}
	The terms on the right-hand side, $\rho_{_E},{\bf j}_{_E}$ and $\rho_{_B},{\bf j}_{_B}$, represent effective charges and currents induced by the gravitational-wave background, which vanish in a fixed Minkowski spacetime. The effective currents are given by
	\begin{align}
		\mathbf{j}_{E} =
		-\frac{1}{2}\Big[ E_x \dot h_+ + E_y \dot h_\times & - B_y \dot h_+ + B_x \dot h_\times, ~ -E_y \dot h_+ - B_x\dot h_+ \nonumber \\
		& + E_x \dot h_\times - B_y \dot h_\times , ~0
		\Big]
		\label{eq:j_e}
		\\
		\mathbf{j}_{B}  =-\frac{1}{2}\Big[ B_x \dot h_+ + B_y \dot h_\times & + E_y \dot h_+ - E_x \dot h_\times, ~ E_x \dot h_+ - B_y \dot h_+ \nonumber \\
		& + B_x \dot h_\times + E_y \dot h_\times , ~0 \Big].
		\label{eq:j_b}
	\end{align}
	Similarly, the fluid equations are given by
	\begin{align}
		\mathbf{e}_0(\gamma n) + \bar{\nabla}\cdot(\gamma n \mathbf{v}) = \Delta n \, ,
		\label{eq:num_axial_1}
	\end{align}
	and
	\begin{align}
		(\epsilon + p)(\mathbf{e}_0 + \mathbf{v}\cdot \bar{\nabla})(\gamma \mathbf{v}) 
		&= -\gamma^{-1}\bar{\nabla}p - \gamma \mathbf{v}(\mathbf{e}_0 + \mathbf{v}\cdot \bar{\nabla})p \nonumber \\
		& + (\epsilon + p)\mathbf{G} + \rho \mathbf{E} + \mathbf{j} \times \mathbf{B} \, .
		\label{eq:num_axial_2}
	\end{align}
	 Charge conservation implies $\Delta n = 0$ and 
	\begin{equation}
		\mathbf{G}^b \equiv -\,\omega^b{}_{ac}\,u^a u^c \, ,
		\label{eq:G-def}
	\end{equation}
	which, for $u^a = \gamma(1, v_x, v_y, v_z)$, evaluates explicitly to
	\begin{align}
		\mathbf{G} = -\frac{\gamma}{2} \Big[
		& v_x \dot{h}_+ - v_x v_z \dot{h}_+ + v_y \dot{h}_\times - v_y v_z \dot{h}_\times, \nonumber \\
		& -v_y \dot{h}_+ + v_y v_z \dot{h}_+ + v_x \dot{h}_\times - v_x v_z \dot{h}_\times, \nonumber \\
		& (v_x^2 - v_y^2) \dot{h}_+ + 2 v_x v_y \dot{h}_\times
		\Big] \, .
		\label{eq:G-def1} 
	\end{align}
	Finally, the chiral anomaly equation, Eq.~\eqref{eq:conserv-1}, takes the form
	\begin{align}
		(\mathbf{e}_0 + \mathbf{v}\cdot \bar{\nabla})(\gamma n_5) + \bar{\nabla}\cdot \mathbf{j}_5 
		= - C \, \mathbf{E}\cdot \mathbf{B} \, .
	\end{align}
	%
	\section{Interaction between gravitational and plasma waves}
	\label{sec-GW-wave-inter}
	Assuming no prior interaction between gravitational and plasma waves, we model a static, homogeneous plasma at rest in flat Minkowski spacetime using Cartesian coordinates. We treat gravitational waves as minor perturbations of the metric, while magnetohydrodynamic (MHD) waves are modelled as small fluctuations around the plasma's equilibrium. Under weak nonlinear coupling, three-wave resonances drive the primary interactions, {\it i.e.}
	\[
	{\bf k}_1 = {\bf k}_2 + {\bf k}_3\, , \qquad f_1 = f_2 + f_3\, ,
	\]
	with ${\bf k}$ a wave vector and $f=f({\bf k})$ the corresponding frequency, fixed by a dispersion relation. In this section, we derive the resonant three-wave interaction between a GW and the MHD waves supported by a chiral plasma, for propagation along the background magnetic field.
	%
	\subsection{Master equations}
	\label{sec-corrected-master}
	For this analysis, we consider a configuration in which the background magnetic field ${\bf B}_0 = B_0 \hat{z}$ is aligned with the propagation direction, ensuring that the background field itself requires no frame correction. We assume purely transverse velocity perturbations, $\delta{\bf v} = \delta v_x \hat{x} + \delta v_y \hat{y}$ (where $\delta v_z = 0$), within a neutral plasma ($n=0$, such that $\rho{\bf E}=0$). Under these conditions, the pressure-gradient terms vanish identically. Furthermore, in the weak-coupling limit, all perturbations depend exclusively on the phase variable $z-t$. Consequently, the transverse tetrad basis vectors $e_1$ and $e_2$ reduce to the coordinate derivatives $\partial_x$ and $\partial_y$, whose action on any such perturbation function vanishes.
    
    The governing equations for the coupled GW-ChMHD dynamics are constructed by synthesizing two complementary components. The first comprises the standard flat-spacetime ChMHD relations, specifically Ohm's, Ampere's, and Faraday's laws, supplemented by the chiral current $\delta{\bf j} = \xi_0\delta\bm\omega + \xi_B^{(0)}\delta{\bf B}$. In the absence of gravitational perturbations, this sector correctly recovers the chiral Alfv\'en waves derived by Yamamoto~\cite{yamamoto2015chiral}. The second component introduces the geometric contributions induced by the GW background, namely the effective source terms ${\bf G}$, ${\bf j}_E$, and ${\bf j}_B$, alongside the GW-generated vorticity $\delta\bm\omega$ (the detailed derivation of which is deferred to Appendix~\ref{app:coupling-channels}). Combining these standard and geometric elements forms a closed system of equations that fully describes the interaction between the gravitational waves and the chiral plasma.
    \begin{enumerate}[label=(\alph*)]
        \item \textbf{Euler equation:} The Lorentz force associated with the chiral current can be written as
        \[
            {\bf j}\times{\bf B} = {\bf B}_0\times \left( \bar\nabla\times\delta{\bf B} - \xi_0\delta\bm\omega \right).
        \]
        Retaining only terms linear in the perturbations, the momentum equation becomes
        \begin{eqnarray}
            (\epsilon_0+p_0)\,\partial_t\delta{\bf v} &=& -{\bf B}_0\times\bar\nabla\times\delta{\bf B}
            +\xi_0\,{\bf B}_0\times\delta\bm\omega \nonumber\\
            &&
            -(\epsilon_0+p_0)\,{\bf G}+{\bf B}_0\times{\bf j}_E, \label{eq:euler-corrected}
        \end{eqnarray}
        where ${\bf G}$ is given by Eq.~\eqref{eq:G-def1} evaluated for $v_z=0$, ${\bf j}_E$ is defined in Eq.~\eqref{eq:j_e}, and
        \[
            \delta\bm\omega = \bar\nabla\times\delta{\bf v} +\delta\bm\omega_{\rm(GW)}
        \]
        includes both the fluid and the GW-induced connection contributions to the vorticity. Consequently, the gravitational wave couples to the Euler equation through three distinct channels: the effective force ${\bf G}$, the induced current ${\bf j}_E$, and the GW-generated contribution to the vorticity entering the chiral vortical effect (CVE).
        \item \textbf{Ohm's law and the induction equation:} In the ideal-MHD limit, Ohm's law,
        \[
            \delta{\bf E} = -\delta{\bf v}\times{\bf B}_0,
        \]
        determines the perturbed electric field appearing in ${\bf j}_E$ and ${\bf j}_B$. Substituting this relation into Faraday's law, Eq.~\eqref{eq:max4}, provides the induction equation
        \begin{equation}
            \partial_t\delta{\bf B} = \bar\nabla\times (\delta{\bf v}\times{\bf B}_0) + {\bf j}_B,
            \label{eq:induction-corrected}
        \end{equation}
        where ${\bf j}_B$ is defined by Eq.~\eqref{eq:j_b}.
        \item \textbf{Closure of the perturbation equations:} In the absence of the GW-induced source terms (${\bf j}_E={\bf j}_B=0$), differentiating Eq.~\eqref{eq:euler-corrected} with respect to time and using Eq.~\eqref{eq:induction-corrected} to eliminate $\partial_t\delta{\bf B}$ leads to a closed second-order equation for the velocity perturbation $\delta{\bf v}$. Once the GW-induced currents ${\bf j}_E$ and ${\bf j}_B$ are retained, however, the elimination is no longer complete. Since ${\bf j}_E$ depends explicitly on $\delta{\bf B}$ rather than solely on its time derivative, the resulting equations retain undifferentiated magnetic-field perturbations. Consequently, the dynamics can no longer be expressed solely in terms of $\delta{\bf v}$. Instead, the coupled GW-ChMHD system must be treated as a first-order system evolving both $\delta{\bf v}$ and $\delta{\bf B}$ simultaneously. This qualitative change in the mathematical structure arises directly from the GW-induced current channels ${\bf j}_E$ and ${\bf j}_B$.
    \end{enumerate}
    \noindent To simplify the governing equations, we introduce
	\begin{eqnarray}
		v_A\equiv \frac{B_0}{\sqrt{\epsilon_0+p_0}}, \quad f_A\equiv v_Ak_z, \quad v_T\equiv\frac{\xi_0v_A}{\sqrt{\epsilon_0+p_0}}~. \nonumber 
	\end{eqnarray}
    The plasma perturbations are Fourier decomposed along the background magnetic field, 
    \[ 
    \delta v_{x,y}\rightarrow \delta v_{x,y}(t)e^{ik_zz}, \qquad \delta B_{x,y}\rightarrow \delta B_{x,y}(t)e^{ik_zz}, 
    \] 
    while the GW amplitudes retain their explicit dependence on both space and time. We also introduce the circularly polarized combinations 
     \begin{align*}
         a_{\rm I}=\delta v_x+i\delta v_y, \qquad a_{\rm II}=\delta v_x-i\delta v_y,
     \end{align*}
    together with the normalized magnetic-field perturbations 
    \begin{align*}
        d_{\rm I} =\frac{\delta B_x+i\delta B_y}{\sqrt{\epsilon_0+p_0}}, \qquad d_{\rm II} =\frac{\delta B_x-i\delta B_y}{\sqrt{\epsilon_0+p_0}},
    \end{align*}
    so that $d_{\rm I}$ and $d_{\rm II}$ have the dimensions of velocity, consistent with the normalization defined by $v_A$. The GW-induced source terms ${\bf G}$, ${\bf j}_E$, and ${\bf j}_B$ contain contributions proportional to both $\dot h_+$ and $\dot h_\times$, with coupling strengths of the same order. Accordingly, neglecting one polarization is not justified on perturbative grounds. Throughout this work, we set $h_+=0$ and restrict the analysis to the cross-polarized GW mode. We adopt this choice solely to reduce the complexity of the coupled GW-ChMHD system and should regard it as a limitation of this analysis. By contrast, Ref.~\cite{servin2003resonant} considered the complementary case ($h_\times=0$), where the plus polarization provides the dominant coupling in the resonance configuration studied there. Whether a comparable simplification exists for the present configuration, or whether both GW polarizations must be retained simultaneously, remains an open question and is left for future analysis.
	
    Substituting the ideal-MHD relation $\delta{\bf E}=-\delta{\bf v}\times{\bf B}_0$ into Eqs.~\eqref{eq:j_e}--\eqref{eq:j_b}, together with the assumptions $h_+=0$ and $\delta B_z=0$, and projecting Eqs.~\eqref{eq:euler-corrected} and \eqref{eq:induction-corrected} onto the transverse ($x,y$) components, leads to the following closed first-order system governing the four dynamical variables 
    \begin{align}
    \dot a_I &= -i k_z v_T\, a_I
    + \tfrac{i}{2}(1-v_T-v_A^2)\dot h_\times\, a_{II}
    \nn\\
    &\quad
    + i k_z v_A\, d_I
    - \tfrac{i}{2}v_A\dot h_\times\, d_{II},
    \nn\\
    \dot a_{II} &= -i k_z v_T\, a_{II}
    - \tfrac{i}{2}(1 - v_A^2- v_T)\dot h_\times\, a_I
    \nn\\
    &\quad
    + i k_z v_A\, d_{II}
    + \tfrac{i}{2}v_A\dot h_\times\, d_I,
    \nn\\
    \dot d_I &= i k_z v_A\, a_I
    -\tfrac{i}{2}v_A\dot h_\times\, a_{II}
    -\tfrac{i}{2}\dot h_\times\, d_{II},
    \nn\\
    \dot d_{II} &= i k_z v_A\, a_{II}
    +\tfrac{i}{2}v_A\dot h_\times\, a_I
    +\tfrac{i}{2}\dot h_\times\, d_I~~.
    \label{eq:corrected-master}
    \end{align}
    Note that, in the absence of gravitational-wave perturbations ($\dot h_\times=0$), all GW-induced couplings vanish identically and Eq.~\eqref{eq:corrected-master} reduces to the uncoupled ChMHD system. The corresponding dispersion relation is therefore recovered exactly,
    \[
        f^2-v_Tk_zf-f_A^2=0,
    \]
    in agreement with the standard chiral Alfv\'en -wave result.

    Equation~\eqref{eq:corrected-master} exhibits two important structural differences compared with formulations that neglect the GW-induced current terms ${\bf j}_E$, ${\bf j}_B$, and the GW-generated contribution to the vorticity. First, the velocity perturbations $a_{\rm I}$ and $a_{\rm II}$ no longer form a closed subsystem. Owing to the explicit dependence of ${\bf j}_B$ on the GW perturbation, the magnetic-field variables $d_{\rm I}$ and $d_{\rm II}$ become independent dynamical degrees of freedom, and the coupled GW-ChMHD equations must therefore be treated as a first-order system for all four variables. This contrasts with the conventional formulation, in which the magnetic-field perturbations can be eliminated to obtain a closed second-order equation for the velocity perturbations.
    Second, the GW-induced couplings generate direct mixing not only between the two circular polarization states ($a_{\rm I}\leftrightarrow a_{\rm II}$), but also between the velocity and magnetic-field perturbations ($a\leftrightarrow d$) at $\mathcal{O}(\dot h_\times)$. Consequently, the GW mediates a simultaneous coupling between the velocity and magnetic-field polarization modes, rather than acting independently on each sector. Both structural modifications arise from the GW-induced current channels ${\bf j}_E$ and ${\bf j}_B$. If these terms are omitted while retaining the effective force ${\bf G}$ and the GW-induced vorticity contribution, the system again reduces to a closed two-variable description.
    %
    \subsection{CVE-only limit}
    \label{sec-cve-only-corrected}
    When the magnetic-tension contribution associated with the ordinary Alfv\'en dynamics is subdominant to the chiral-vortical convection, the four-variable system introduced above admits a reduced two-variable description that isolates the CVE-driven dynamics. This reduction follows directly from Eq.~\eqref{eq:corrected-master} in the corresponding limiting regime.
	
    In the set of eqs.~\eqref{eq:corrected-master}, $d_I,d_{II}$ enter the $a_I,a_{II}$ equations only through the magnetic-tension term $ik_zv_Ad_I$ and the ${\bf j}_E$-driven term $\tfrac{i}{2}v_A\dot h_\times d_{II}$, while the $a_I\leftrightarrow a_{II}$ cross-coupling coefficient $\tfrac{i}{2}(1-v_T-v_A^2)$ combines a $v_A$ independent term from ${\bf G}$ with $v_T$ and $v_A^2$ terms that arises from the GW-vorticity and ${\bf j}_E$ channels respectively. Physically, $v_A$ measures the strength of ordinary Alfv\'en-wave inertia relative to the CVE convection $v_T$. In the regime where $v_T\gg v_A$, every $v_A$-proportional term in eqs.~\eqref{eq:corrected-master} is subleading and can consistently be dropped together. Consequently, $d_I,d_{II}$ decouple entirely from the $a_I, a_{II}$ equations. Writing $A_I, A_{II}$ for $a_I, a_{II}$ in this reduced system, to keep the two treatments notationally distinct, we obtain
    \begin{eqnarray}
        \dot A_I &=& -ik_zv_T\,A_I \,+\,\tfrac{i}{2}(1-v_T)\dot h_\times\,A_{II}\, , \label{eq:cve-only-AI}\\
        \dot A_{II} &=& -ik_zv_T\,A_{II} \,-\,\tfrac{i}{2}(1-v_T)\dot h_\times\,A_I\, . \label{eq:cve-only-AII}
    \end{eqnarray}
    It is worth noting that the formal limit $v_A\rightarrow 0$ does not generate a self-coupling term proportional to $\dot h_\times A_{\rm I}$ in the equation for $\dot A_{\rm I}$, nor the corresponding term in the equation for $\dot A_{\rm II}$. Instead, the GW contribution from ${\bf G}$ enters exclusively through the cross-polarization coupling between $A_{\rm I}$ and $A_{\rm II}$. This reduced system, therefore, provides both a useful description of the CVE-dominated regime and an analytical consistency check on the full numerical treatment presented in the forthcoming section.
    
    Alternatively, the same set of equations can be recovered by deriving them directly from the momentum equation after neglecting the magnetic-tension term from the outset,
    \begin{align}
        ~~~~(\epsilon_0+p_0)\,\partial_t\delta{\bf v} = \xi_0\,({\bf B}_0\times\delta\bm\omega) - (\epsilon_0+p_0)\,{\bf G}\, ,
        \label{eq:cve-only-start}
    \end{align}
    where $\delta\bm\omega=\bar{\nabla}\times\delta{\bf v}+\delta\bm\omega_{\rm (GW)}$ as before. The current ${\bf j}_E$ is also omitted, together with the magnetic-tension term, because both arise from the same physical channel through the generalized Amp\`ere's law, eq.~\eqref{eq:max3}. Retaining ${\bf j}_E$ while neglecting the magnetic-tension term would reintroduce a dependence on $\delta{\bf B}$ without a corresponding evolution equation to determine $\delta{\bf B}$.
    \begin{align}
        \partial_t\,\delta v_x &= -v_T\,\partial_z\,\delta v_x \,+\, \tfrac12(1-v_T)\,\dot h_\times\,\delta v_y\, , \\
        \partial_t\,\delta v_y &= -v_T\,\partial_z\,\delta v_y \,+\, \tfrac12(1-v_T)\,\dot h_\times\,\delta v_x\, .
    \end{align}
    Combining the above equations by defining $A_I=\delta v_x+i\delta v_y$ and $A_{II}=\delta v_x-i\delta v_y$ gives exactly eqs.~\eqref{eq:cve-only-AI}-\eqref{eq:cve-only-AII}. Thus, the agreement between the two independent derivations provides another consistency check on the underlying linearised analysis.
    %
	\subsection{Slowly-Varying-Envelope analysis (SVEA) at exact resonance}
	\label{sec-svea-corrected}
	We derive the resonant growth rate $\Gamma$ at exact resonance analytically, through SVEA applied to eqs.~\eqref{eq:corrected-master}. This is a special case of the Floquet treatment, which covers arbitrary drive frequency $f_g$ without envelope truncation. We compare the two results in the later section.
	
	At $\dot h_\times=0$, eqs.~\eqref{eq:corrected-master} decouple into two doublets, $(a_{I_0},d_{I_0})$ and $(a_{II_0},d_{II_0})$. The four-variable system splits exactly into two identical decoupled $2\times2$ doublets,
	\begin{align}
		\dot a_{I_0} &= -ik_zv_T\,a_{I_0} + ik_zv_A\,d_{I_0} , \quad
		\dot d_{I_0} = ik_zv_A\,a_{I_0} , \label{1a}\\
		\dot a_{II_0} &= -ik_zv_T\,a_{II_0} + ik_zv_A\,d_{II_0} , \quad
		\dot d_{II_0} = ik_zv_A\,a_{II_0} . \label{1b}
	\end{align}
	Since both doublets are governed by the same matrix, it suffices to solve the $(a_{I_0},d_{I_0})$ system. The solution of the $(a_{II_0},d_{II_0})$ doublet follows identically. With $\mathbf X=(a_{I_0},d_{I_0})^T$, Eq.~\eqref{1a} can be written as $\dot{\mathbf X}=M_0\mathbf X$ where
    \begin{equation}
	M_0 =
		\begin{pmatrix}
			-ik_zv_T & ik_zv_A \\
			ik_zv_A & 0
		\end{pmatrix}.
	\label{eq:M0}
    \end{equation}
    Assume solutions of the form $\mathbf X(t) = \mathbf X_0\, e^{-if t}$. Substituting into $\dot{\mathbf X}=M_0\mathbf X$ gives 
    \begin{equation*}
	-if\,\mathbf X_0 = M_0\,\mathbf X_0
	\qquad\Longleftrightarrow\qquad
	(iM_0)\,\mathbf X_0 = f\, \mathbf X_0 .
    \end{equation*}
    Define the real matrix
    \begin{equation*}
	N \equiv i M_0 =
	\begin{pmatrix}
		k_zv_T & -k_zv_A \\
		-k_zv_A & 0
	\end{pmatrix},
	\label{eq:N}
    \end{equation*}
    so that the problem becomes the ordinary eigenvalue equation
    \begin{equation*}
	N\,\mathbf X_0 = f\,\mathbf X_0 .
	\label{eq:eigprob}
    \end{equation*}
    The characteristic equation, $\det(N- f\mathbf I)=0$ where $\mathbf I$ is the identity matrix, is given by
    \begin{equation*}
	f^2 - v_Tk_z\,f - (v_Ak_z)^2 = 0 .
    \end{equation*}
    With $f_A\equiv v_Ak_z$, the above equation reduces exactly to the free chiral-Alfv\'en dispersion relation
    \begin{equation*}
	f^2 - v_Tk_z\,f - f_A^2 = 0 .
	\label{eq:disprel}
    \end{equation*}
    The eigenvalues are,
    \begin{equation*}
	f_\pm = \frac{v_Tk_z \pm \sqrt{(v_Tk_z)^2 + 4f_A^2}}{2}
	\label{eq:eigenvalues}
    \end{equation*}
    with
    \begin{equation*}
        f_+ + f_- = v_Tk_z = \mathrm{tr}\,N,
	\quad
	f_+\,f_- = -f_A^2 = \det N .
    \end{equation*}
    Both roots are real for real $v_T,v_A,k_z$ (since the discriminant $(v_Tk_z)^2+4f_A^2\ge0$ always), and satisfy $f_+>0>f_-$ when $f_A\neq0$. The eigenvectors corresponding to the eigenvalues $f_\pm$  are
    \begin{equation*}
	\mathbf X_0(f_\pm) = (a_I,d_I) \propto
	\left(1,\ -\frac{f_A}{f_\pm}\right)~.
	\label{eq:eigenvectors}
    \end{equation*}
    We expand the driven solution in this free eigenbasis, with slowly varying coefficients $P, Q, R, S$,
    \begin{align}
	a_I &\approx P e^{-if_+t} + Q e^{-if_-t}\, ,\nn \\
	d_I & \approx -\tfrac{f_A}{f_+}P e^{-if_+t} - \tfrac{f_A}{f_-}Q e^{-if_-t}\, , \nn\\
	a_{II} &\approx R e^{-if_+t} + S e^{-if_-t}\, , \nn \\
	d_{II} &\approx -\tfrac{f_A}{f_+}R e^{-if_+t} - \tfrac{f_A}{f_-}S e^{-if_-t}\, ,
	\label{eq:SVEA_1}
    \end{align}
    The envelope approximation enters only when the resulting equations for $P, Q, R, S$ are expanded, and the fast-oscillating terms are dropped.
    Writing 
    \begin{align*}
	\begin{pmatrix}
		a_I \\
		d_I
	\end{pmatrix} = T
	\begin{pmatrix}
		P \\
		Q
	\end{pmatrix}
    \end{align*}
    with
    \begin{align*}
	T(t) &=
	\begin{pmatrix}
		e^{-if_+t} & e^{-if_-t} \\[2pt]
		-\dfrac{f_A}{f_+}e^{-if_+t} &
		-\dfrac{f_A}{f_-}e^{-if_-t}
	\end{pmatrix}. 
    \end{align*}
    The columns of $T$ are the eigenvectors of the free evolution operator $M_0$.
	
    \noindent
    Decomposing  
    \[
    	\frac{d}{dt} \begin{pmatrix}
    		a_I\\
    		d_I
    	\end{pmatrix} 
        = M_0\begin{pmatrix}
    		a_I \\
    		d_I
    	\end{pmatrix} + \begin{pmatrix}
    		\lambda_{a_I}\\\lambda_{d_I}
    	\end{pmatrix}\, 
	\]
    where $\lambda_{a_I,} \lambda_{d_I}$ are the coupling terms given as
    \begin{align}
    	\lambda_{a_I} &= \tfrac{i}{2}(1-v_T-v_A^2)\dot h_\times a_{II} - \tfrac{i}{2}v_A\dot h_\times d_{II}\, , \label{eq:lambda_aI}\\
    	\lambda_{d_I} &= -\tfrac{i}{2}v_A\dot h_\times a_{II} - \tfrac{i}{2}\dot h_\times d_{II}\, \label{eq:lambda_dI}.
    \end{align}
    Using $\dot T=M_0~T$, the free parts cancel exactly, leaving
    \begin{equation}
    	\frac{d}{dt} 
        \begin{pmatrix}
    		P\\
    		Q
    	\end{pmatrix} = T^{-1}(t)\times
        \begin{pmatrix}
    		\lambda_{a_I}\\\lambda_{d_I}
    	\end{pmatrix}\, ,
        \label{eq:PQdot-exact}
    \end{equation}
    A similar expression can be obtained for $(R, S)$ with $a_I,d_I$ in place of $a_{II},d_{II}$ in the eigenvalues and an overall sign flip. 

    Substituting the free-mode expansion of $a_{II}$ and $d_{II}$ from eq.~\eqref{eq:SVEA_1} in the coupling term eq.~\eqref{eq:lambda_aI}-eq.~\eqref{eq:lambda_dI} we obtain
    \begin{align}
    	\lambda_{a_I} &=\dot h_\times\Big[c_1 R\,e^{-if_+t}+c_2 S\,e^{-if_-t}\Big], \label{eq:lambda_aI-freq}\\
	   \lambda_{d_I} &=\dot h_\times\Big[c_3 R\,e^{-if_+t}+c_4 S\,e^{-if_-t}\Big],
	   \label{eq:lambda_dI-freq}
    \end{align}
    where constants $c_1,c_2,c_3,c_4$ are constructed from $v_A,v_T,f_A,f_\pm$. Assuming $h_\times = h_0 \cos(f_g~t)$, substituting its time derivative in eq.~\eqref{eq:lambda_aI-freq} and eq.~\eqref{eq:lambda_dI-freq}, and multiplying by $T^{-1}(t)$ decomposes each of $\lambda_{a_I},\lambda_{d_I}$  into four exponential components, one for each choice of $f\to+f_g$ or $-f_g$ combined with each free-mode frequency $-f_+$ or $-f_-$, {\it i.e.}
	\begin{align}
		\lambda_{a_I},\lambda_{d_I}
		&\rightarrow \Big\{
		e^{i(f_g-f_+)t},
		e^{i(f_g-f_-)t}, \nn \\ 
		&\qquad
		e^{i(-f_g-f_+)t},
		e^{i(-f_g-f_-)t}
		\Big\}.
		\label{eq:four-terms}
	\end{align}
	From Eq.~(\ref{eq:PQdot-exact}), we obtain
	\begin{align}
		\dot P &= \frac{f_A\,e^{if_+t}}{f_+-f_-}
		\Big[-\tfrac{f_A}{f_-}\,\lambda_{a_I} - \lambda_{d_I}\Big] \\
		\dot Q &= \frac{f_A\,e^{if_-t}}{f_+-f_-}
		\Big[~\tfrac{f_A}{f_+}\,\lambda_{a_I} + \lambda_{d_I}\Big] 
	\end{align}
	Multiplying the prefactor $e^{if_+t}$ (or $e^{if_-t}$ ) into each of the four terms of Eq.~\eqref{eq:four-terms} shifts every exponent by $+f_+$
	(or $+f_-$). For instance, 
	\begin{align*}
		\dot P  \supset\ e^{i\big(f+f_+\big)t}, \quad
		\dot Q \supset\ e^{i\big(f+f_-\big)t}
	\end{align*}
	where,
	\begin{align}
		~~~~f \in \{f_g, ~f_g+f_+-f_-, ~ -f_g, ~ -f_g+f_+-f_- \}.
		\label{eq:four-exponents}
	\end{align}
	A term is secular, i.e.\ produces steady growth of the slowly varying envelope $P,Q$ rather than a fast oscillation that time-averages to zero under the standard multiple-scales/SVEA argument,  when its exponent in Eq.~\eqref{eq:four-exponents} vanishes. The four shifted frequencies are examined for the secularity condition, and they are listed below.
	\begin{itemize}
		\item $\pm f_g=0$ is a trivial case with no drive and hence, excluded.
		\item $f_g+f_+-f_-=0 \ \Rightarrow\ f_g=f_--f_+$. Since $f_+>0>f_-$ $\ \Rightarrow\ f_--f_+<0$. Since the drive frequency $f_g$ is by convention positive, this scenario is also excluded. 
		\item $-f_g+f_+-f_-=0 \ \Rightarrow\ f_g=f_+-f_-$ is the only physically admissible, nonzero, positive drive frequency.
	\end{itemize}
	The secular condition can also be given as,
	\begin{equation}
		f_g=f_+-f_-\equiv|f_+|+|f_-|\, ,
		\label{eq:resonance-condition}
	\end{equation}
	the sum-frequency resonance condition. Since $\lambda_{a_I},\lambda_{d_I}$ have no $a_I,d_I$ dependence, $\dot P$ picks up no $Q$-dependent secular term at all. The same calculation for $\dot Q,\dot R,\dot S$ follows by the $I\leftrightarrow II$ symmetry of eqs.~\eqref{eq:corrected-master}. At exact resonance,
	\begin{align}
		\dot P &= M_{PS}\,S\, , \qquad \dot S = M_{SP}\,P\, ,\nn\\
		\dot Q &= M_{QR}\,R\, , \qquad \dot R = M_{RQ}\,Q\, , \label{eq:corrected-envelope}
	\end{align}
	with
	\begin{eqnarray}
		M_{PS} &=& \frac{f_Ah_0}{4}\left[\frac{f_A^2}{f_-^2}v_A-\frac{f_A}{f_-}(v_A^2+v_T-2)-v_A\right]\, , \nn\\
		M_{SP} &=& -\frac{f_Ah_0}{4}\left[\frac{f_A^2}{f_+^2}v_A-\frac{f_A}{f_+}(v_A^2+v_T-2)-v_A\right]\, , \nn
	\end{eqnarray}
	and $(M_{QR},M_{RQ})$ obtained from $(M_{PS},M_{SP})$ by $f_+\leftrightarrow f_-$. The $4\times4$ system therefore block-diagonalizes into two independent $2\times2$ blocks, $(P,S)$ and $(Q,R)$.
    \begin{figure}[!t]
        \centering
        \includegraphics[width=0.35\textwidth,height=0.35\textwidth]{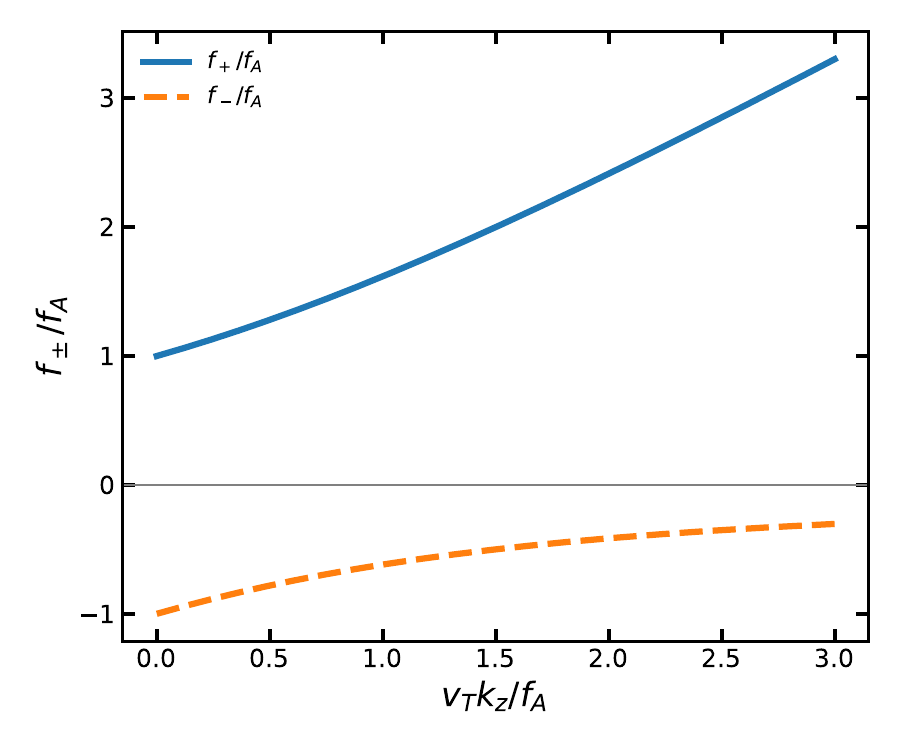}
        \caption{\raggedright The normalized free-dispersion branches, $f_\pm/f_A$, as a function of the CVE-to-Alfv\'en strength ratio, $v_Tk_z/f_A$. The branches are degenerate at $v_Tk_z=0$, and exhibit progressive splitting as the chiral vortical effect (CVE) increases.}
        \label{fig:1a}
    \end{figure}
    
    Using $f_+f_-=-f_A^2$, $f_++f_-=v_Tk_z$, the product $M_{PS}M_{SP}$ simplifies to a perfect square,
    \begin{equation}
        M_{PS}M_{SP}=\frac{k_z^2v_A^2h_0^2}{16}\big(v_A^2+2v_T-2\big)^2\, .
    \end{equation}
    Since the product $M_{\text{PS}}M_{\text{SP}}$ is a perfect square and thus strictly non-negative, the block's eigenvalues $\lambda = \pm\sqrt{M_{\text{PS}}M_{\text{SP}}} = \pm\Gamma$ are purely real. This eigenvalue pair gives a solution consisting solely of growing ($e^{+\Gamma t}$) and decaying ($e^{-\Gamma t}$) exponential modes. Consequently, exact resonance produces purely exponential growth, unlike the modulated envelopes characteristic of complex eigenvalues or the bounded oscillations that arise off-resonance when $M_{\text{PS}}M_{\text{SP}}<0$. With $f_A=v_Ak_z$, we obtain an analytic form of growth rate as
    \begin{equation}
        \Gamma=\frac{f_Ah_0}{4}\big|v_A^2+2v_T-2\big|=\frac{h_0f_A}{2}\Big|1-\frac{v_A^2}{2}-v_T\Big|\, ,
        \label{eq:envelope-matrix}
    \end{equation}
    An analogous result is obtained for the $(Q,R)$ block as well. 
    
    For the fiducial values of the parameters $f_A=1$, $v_A=1$, $k_z=1$ , 
    and $h_0=0.05$, Eq.~\eqref{eq:envelope-matrix} gives $\Gamma=0.005000$. Direct numerical integration of Eqs.~\eqref{eq:corrected-master} shows agreement to within three to four significant figures. Throughout this analysis, and in all subsequent numerical results, we adopt a normalization where $k_z=1$. This choice of normalization, along with $c=1$, allows us to express all frequencies in units of $k_z$. Accordingly, $v_A$ and $v_T$ act as dimensionless velocities, while $f_A=v_A k_z$ and $v_T k_z$ represent frequencies. The quoted numerical values (e.g., $v_T k_z=0.3$ and $f_A=1$) should therefore be understood as dimensionless frequencies in these units. We adopt this choice deliberately. Normalizing to $k_z$ rather than the Hubble parameter $H_*$ ensures that the resonance structure, its validation, and its dependence on chirality remain intrinsic properties of the master equation, independent of any specific cosmological epoch. Once a specific epoch is introduced, starting with the self-consistency condition discussed below and continuing thereafter, all relevant quantities are expressed as ratios to $H_*$ (e.g., $f_A/H_*$, $\chi/H_*$, and $\Gamma/H_*$). Only in this form do these parameters acquire direct physical significance within a cosmological setting.

	It is important to reemphasize that Eqs.~\eqref{eq:corrected-master} retain $h$ and $\delta v$ only to the order needed to close a linear system with time-periodic coefficients. A fully nonlinear treatment would require solving the coupled nonlinear chiral-MHD along with linearized gravity. This numerical exercise is left for future work. Moreover, whether additional nonlinear chiral effects exist also remains open. Instead, we report the linear (Floquet) results first, since they are tractable with the equations in hand and have already indicated some interesting results. 
    \begin{figure*}[!ht]
        \centering
        \subfloat[]{\label{fig:2a}
            \includegraphics[width=0.35\textwidth,height=0.35\textwidth]{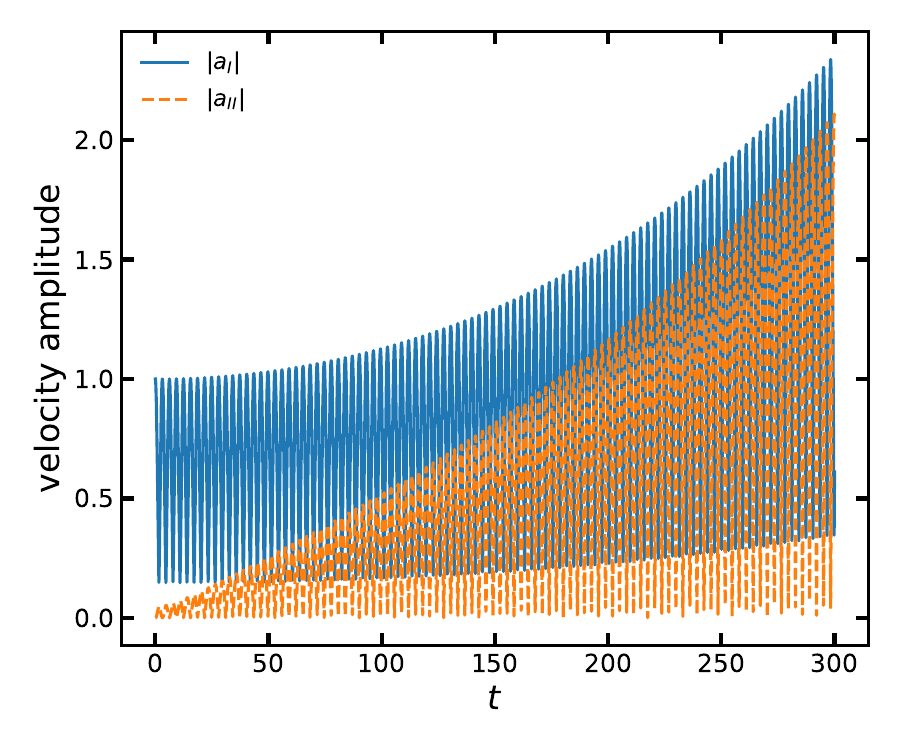}
        }
        \subfloat[]{\label{fig:2b}
            \includegraphics[width=0.35\textwidth,height=0.35\textwidth]{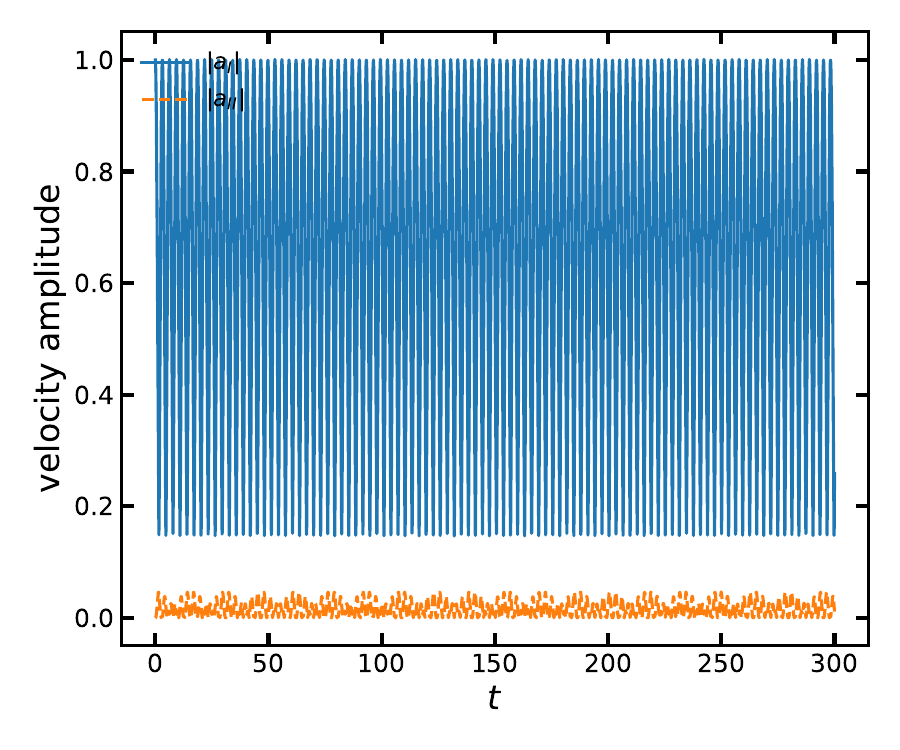}
        }
        
        \vspace{0.2cm} 
        
        \subfloat[]{\label{fig:2c}
            \includegraphics[width=0.35\textwidth,height=0.35\textwidth]{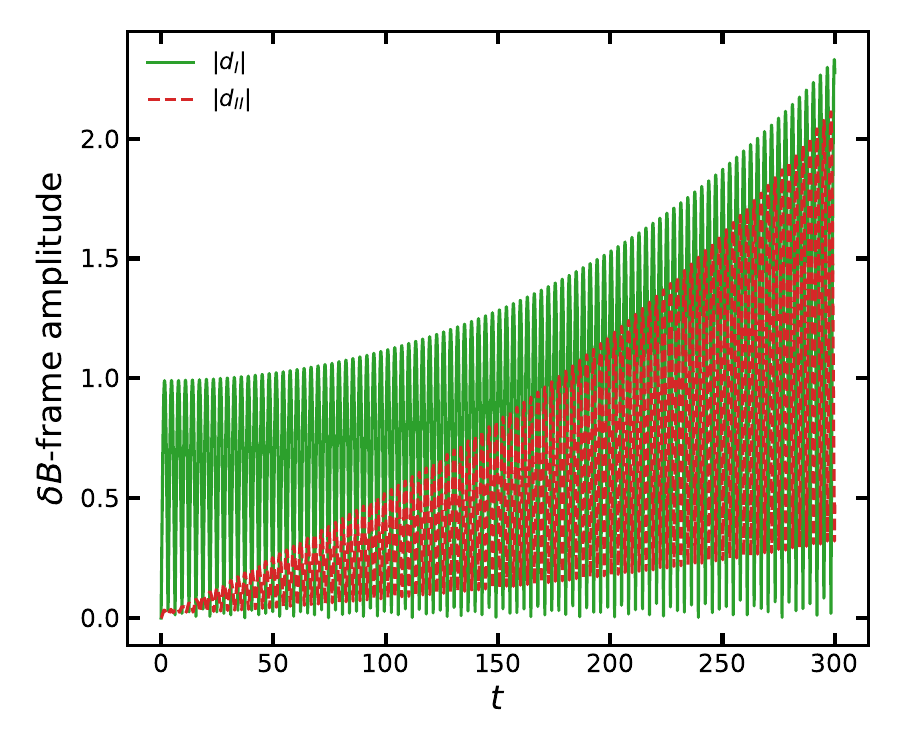}
        }
        \subfloat[]{\label{fig:2d}
            \includegraphics[width=0.35\textwidth,height=0.35\textwidth]{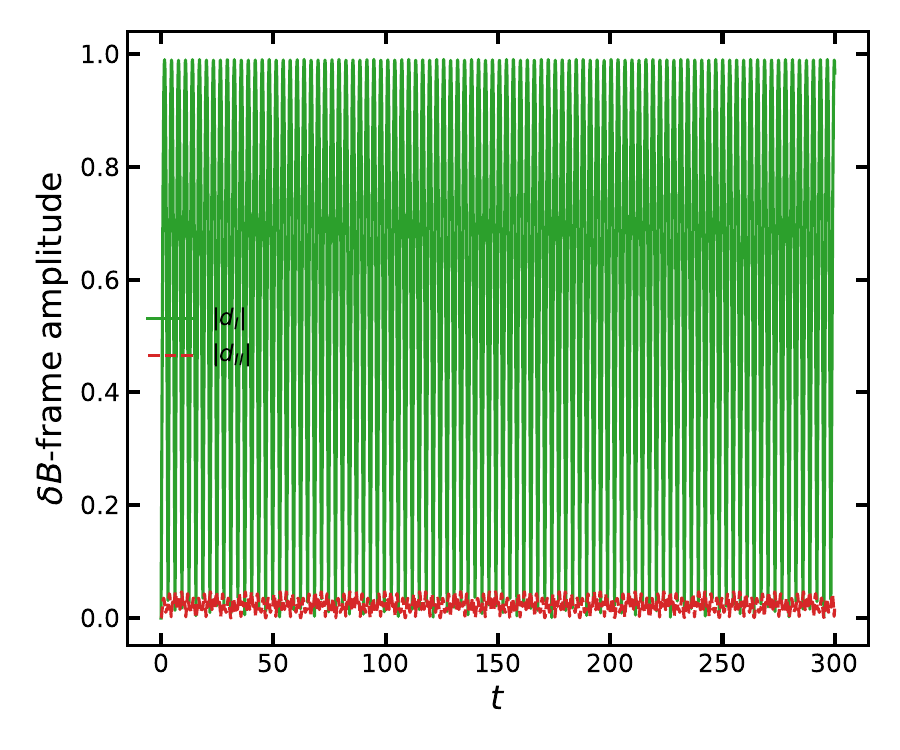}
        }
        
        \caption{\raggedright Time evolution of the velocity amplitudes $a_I$, $a_{II}$ and the corresponding $\delta B$-frame amplitudes $d_I$, $d_{II}$, for $f_A=1$, $v_Tk_z=0.3$, $v_A=1$, $h_0=0.05$ starting from identical initial conditions. (a) The exact resonance case, which exhibits exponential growth for the velocity field. (b) Bounded oscillations for the velocity field detuned by $20\%$. (c) The exact resonance case for the $\delta B$-frame amplitudes $d_I$ and $d_{II}$. (d) Bounded oscillations for the $\delta B$-frame amplitudes detuned by $20\%$.}
        \label{fig:branches-timedomain}
    \end{figure*}
	\subsection{GW Backreaction and Energy Depletion}
	\label{sec-backreaction}
	Distinct from the truncation order discussed above, Eqs.~\eqref{eq:corrected-master} inherit a second limitation by treating $h_\times$ as a fixed external pump in accordance with our overall test-fluid approximation. Linear parametric growth is then unbounded, which cannot be physical. Here, we derive the leading-order correction that accounts for the self-consistent backreaction.
    
	For a background magnetic field $\bm{B}_0 = B_0\hat{z}$ with transverse magnetic perturbations $\delta B_x = \delta F^{23}$ and $\delta B_y = \delta F^{31}$, a direct evaluation of the perturbed electromagnetic stress tensor, $\delta T^{\mu\nu}_{\rm em} = \delta F^{\mu\lambda}F^\nu{}_\lambda + F^{\mu\lambda}\delta F^\nu{}_\lambda - \tfrac12g^{\mu\nu}\delta(F^2)$, reveals that every term vanishes identically. Mathematically, this occurs because the background tensor ($F^{12}$) and the perturbations ($F^{23}, F^{31}$) share no common index pairs. Consequently, for $\bm{B}_0 \parallel \hat{z}$, the electromagnetic stress cannot source gravitational waves. This behavior contrasts sharply with a transverse background field, $\bm{B}_0 \parallel \hat{x}$ (as in Ref.~\cite{kallberg2004nonlinear}), where the electromagnetic stress directly sources the $h_+$ polarization. In our longitudinal configuration, the leading non-vanishing source is instead the $\mathcal{O}(\delta v^2)$ kinetic component of the fluid stress tensor,
	\begin{widetext}
		\beq
		\delta T^{11}-\delta T^{22} \approx (\epsilon_0+p_0)(\delta v_x^2-\delta v_y^2)\, , \qquad
		\delta T^{12}+\delta T^{21} \approx 2(\epsilon_0+p_0)\,\delta v_x\delta v_y\, ,
		\eeq
	\end{widetext}
	so that, using the (already established) wave equation $\Box h_\times=-8\pi G(\delta T_{12}+\delta T_{21})$,
	\be
    	\Box h_\times = -16\pi G(\epsilon_0+p_0)\,\delta v_x\delta v_y\, .
    	\label{eq:backreaction-source}
	\ee
	Since $\delta v_x,\delta v_y\propto e^{ik_zz}$, the product $\delta v_x\delta v_y\propto e^{2ik_zz}$. Momentum conservation requires the sourced GW to sit at $k_g=2k_z$, a degenerate down-conversion in which one GW quantum corresponds to two plasma quanta at the same $k_z$, split in frequency by the CVE term.

	We express the wave as $h_\times(t) = \mathcal{A}(t)e^{-i f_g t} + \overline{\mathcal{A}(t)}e^{i f_g t}$, where the envelope $\mathcal{A}(t)$ is assumed to be slowly varying. This ensures backreaction acts as a small secular correction over a single period, capturing the regime of gradual wave depletion rather than abrupt disruption. Applying this slowly varying envelope approximation to demodulate Eq.~\eqref{eq:backreaction-source} at the resonant wavenumber $k_g = 2k_z$ gives
	\be
    	\dot{\mathcal A} = -\chi\, a_I\,\overline{a_{II}}\, , \qquad
    	\chi \equiv \frac{\pi G(\epsilon_0+p_0)}{f_g} = \frac{H_*^2}{2f_g}\, .
    	\label{eq:pump-depletion}
	\ee

	In the last equality, we have used the radiation-era Friedmann relation $G(\epsilon_0+p_0)=H_*^2/2\pi$. Furthermore, performing the demodulation process with strictly real-valued fields for $\delta v_x, \delta v_y$, and $h_\times$, rather than working directly with complex mode amplitudes, naturally results in the $a_{\text{I}}\bar{a}_{\text{II}}$ structure in the above equation. 
    
    Equation~\eqref{eq:corrected-master} together with eq.~\eqref{eq:pump-depletion} form a closed system in the five variables $(a_I,a_{II},d_I,d_{II},\mathcal A)$. Two features of this closed system can be checked analytically. First, in the undepleted-pump limit ($\chi\to0$), the five-variable system reduces exactly to the (linear) master equations~\eqref{eq:corrected-master}, as it must. Second, for the case where $\chi \neq 0$, the initial rate of pump depletion follows from Eq.~\eqref{eq:pump-depletion} when evaluated at $t=0$. Since the intensity derivative is given by $d\vert{}\mathcal A\vert{}^2/dt = 2\,\mathrm{Re}[\overline{\mathcal A}\,\dot{\mathcal A}]$, substituting $\dot{\mathcal A}(0) = -\chi\,a_I(0)\overline{a_{II}(0)}$ under the uniform daughter-mode initial conditions ($d_I(0) = d_{II}(0) = 0$) provides the initial slope, given by
	\begin{equation}
		\left.
		\frac{d|\mathcal A|^2}{dt}
		\right|_{t=0}
		=
		-2\chi\,\mathrm{Re}\!\left[
		\overline{\mathcal A(0)}\,a_I(0)\overline{a_{II}(0)}
		\right]. 
	\end{equation}
	The slope thus evaluated is unambiguously negative (depletion) for any $\chi>0$, when $a_I(0), a_{II}(0),\mathcal A(0)$ are all taken real and positive. Varying either the relative phase of the daughter modes or the pump's own phase flips this sign since the sign is controlled symmetrically by either phase. Beyond this initial instant, the coupled system must be integrated numerically.  
    \begin{figure*}[t]
        \centering
        \subfloat[]{\label{fig:3a}
            \includegraphics[width=0.35\textwidth,height=0.32\textwidth]{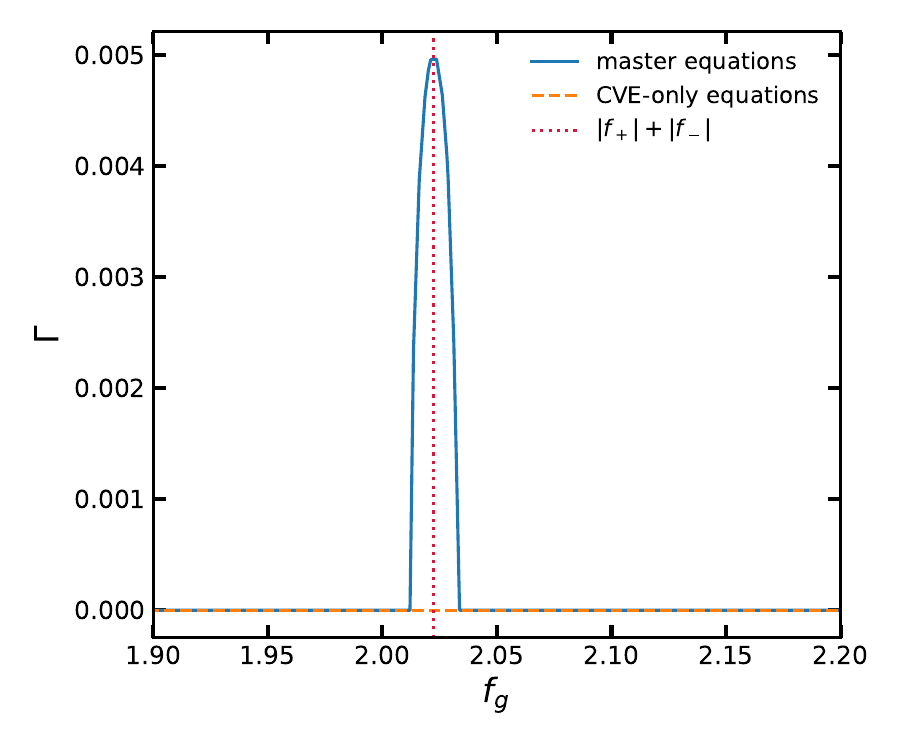}
        }
        \subfloat[]{\label{fig:3b}
            \includegraphics[width=0.45\textwidth,height=0.32\textwidth]{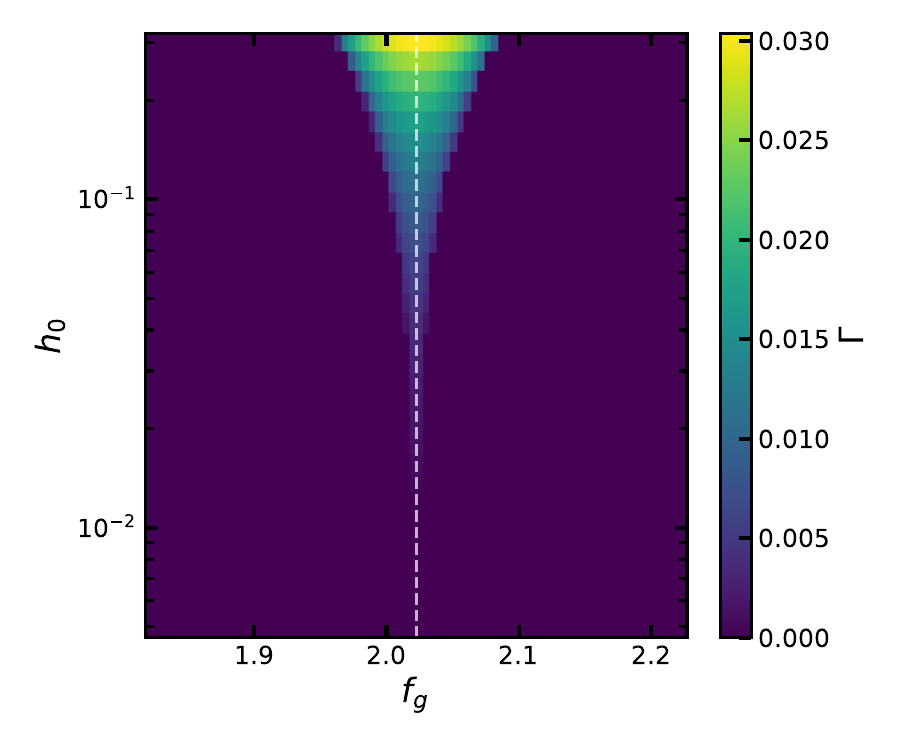}
        }
     
        \caption{\raggedright Floquet growth rate as a function of the gravitational wave frequency $f_g$ for both the full master equations and the CVE-only limit evaluated at $f_A=1$, $v_Tk_z=0.3$, $v_A=1$, and $h_0=0.05$ (in arbitrary units). (a) The parametric resonance occurs exactly at $f_g=\vert{}f_+\vert{}+\vert{}f_-\vert{}$. (b) The growth rate mapped across the $(f_g, h_0)$ parameter space, illustrating the characteristic Arnold-tongue widening as the drive strength increases.}
        \label{fig:resonance-scan}
    \end{figure*}
	An additional self-consistency requirement is that $f_A, f_g \gg H_*$, ensuring many oscillation periods per Hubble time. This condition is necessary to justify the fixed-background approximation employed throughout our analysis. Consequently, the wavenumber $k_z$ is restricted to the deep sub-horizon regime ($k_z \gg H_*$) rather than horizon scales, a constraint that is also directly evident from Eq.~\eqref{eq:pump-depletion}.
    Assuming $a_{\text{I}}, a_{\text{II}} \sim a$ representing the characteristic dimensionless mode-amplitude scale, the rate $\dot{\mathcal{A}} \sim \chi a^2$ is independent of $\mathcal{A}$ itself during the initial-slope regime. Assuming $\vert{}\mathcal{A}(0)\vert{} \sim \mathcal{O}(1)$ in these units, the characteristic depletion timescale is $t_{\text{dep}} \sim \vert{}\mathcal{A}(0)\vert{} / \vert{}\dot{\mathcal{A}}\vert{} \sim 1/(\chi a^2)$. Comparing $t_{\text{dep}}$ with the Hubble time $t_{\text{H}} \sim H_*^{-1}$ gives
    \[
        \frac{t_{\text{dep}}}{t_{\text{H}}} \sim \frac{H_*}{\chi a^2} = \frac{2(f_g/H_*)}{a^2} \approx \frac{4 v_{\text{A}} (k_z / H_*)}
        {a^2}
    \] 
    where $f_g \approx 2f_{\text{A}} = 2v_{\text{A}}k_z$. Applying the fiducial standard $v_{\text{A}} = 1$ this ratio reduces to $t_{\text{dep}}/t_{\text{H}} \sim 4(k_z / H_*) / a^2$. Consequently, for sub-horizon modes ($k_z / H_* \gg 1$) and linear plasma fluctuations ($a \ll 1$), backreaction operates gradually across many oscillation periods rather than abruptly within a single cycle.
    %
	\section{Numerical Solutions and Results}
	\label{sec-results}
	Once the GW strain $h_\times(t)$ is specified, Eq.~\eqref{eq:corrected-master} as well as Eqs.~\eqref{eq:cve-only-AI}--\eqref{eq:cve-only-AII}, in the limit of negligible Alfv\'en-wave inertia, constitute a linear system with time-periodic coefficients. We have integrated it numerically and the main findings are organized around three physical questions- (i) where and how strongly does the resonance occur (Sec.~\ref{sec-res-struct}), (ii) what does it cost the GW pump to drive it (Sec.~\ref{sec-longtime}), and (iii) how does all of this depend on the epoch of the Universe's history at which the source GW and the resonant plasma actually meet (Secs.~\ref{sec-results-benchmark}).
    
	Before interpreting the results below, we validated the implementation through three independent analytic checks. First, for $h_\times=0$ the numerically integrated solution oscillates at frequencies matching the analytic roots of $f^2-v_Tk_z\,f-f_A^2=0$ to within $0.3\%$,  with the residual difference limited by the finite integration time and spectral resolution. For example, for $f_A=1,v_Tk_z=0.3$, the analytic roots are recovered numerically as $ 0.86119$ and $1.16119$. Second, the Floquet growth rate of the undriven ($h_0=0$) vanishes for both the full four-variable system and the CVE-only reduction. This confirms that the $iv_Tk_z\dot a$ term is numerically gyroscopic rather than dissipative, despite its imaginary coefficient. Third, in the limit $v_A=0$ (with $d_I(0)=d_{II}(0)=0$) the full system reduces \emph{exactly} to the CVE-only system. This provides an additional check that the magnetic-tension and Maxwell-sector contributions are correctly confined to the $v_A$-proportional pieces of eq.~\eqref{eq:corrected-master}.
   
	\subsection{Parametric-resonance structure}
    \label{sec-res-struct}
	The free ($h_\times=0$) dispersion relation $f^2-v_Tk_z\,f-f_A^2=0$ has two roots $f_\pm$. In the absence of the CVE term, these reduce to the degenerate values $f_\pm$, while increasing $v_Tk_z$ progressively splits the two branches (Fig.~\ref{fig:1a}). This frequency splitting determines the resonant sum frequency. Direct time integration concretely confirms the resonance. Starting from the same initial condition, the amplitudes $|a_I|,|a_{II}|$ grow exponentially when driven at $f_g=|f_+|+|f_-|$  but remain bounded and oscillatory for a detuned frequency (Fig.\ref{fig:2a} and \ref{fig:2b}).  Similarly behaviour is obtained for $\delta B$-frame amplitudes $|d_I|,|d_{II}|$ as well (Fig.\ref{fig:2c} and \ref{fig:2d}).
	\begin{figure}
		\centering
		\includegraphics[width=0.45\textwidth,height=0.33\textwidth]{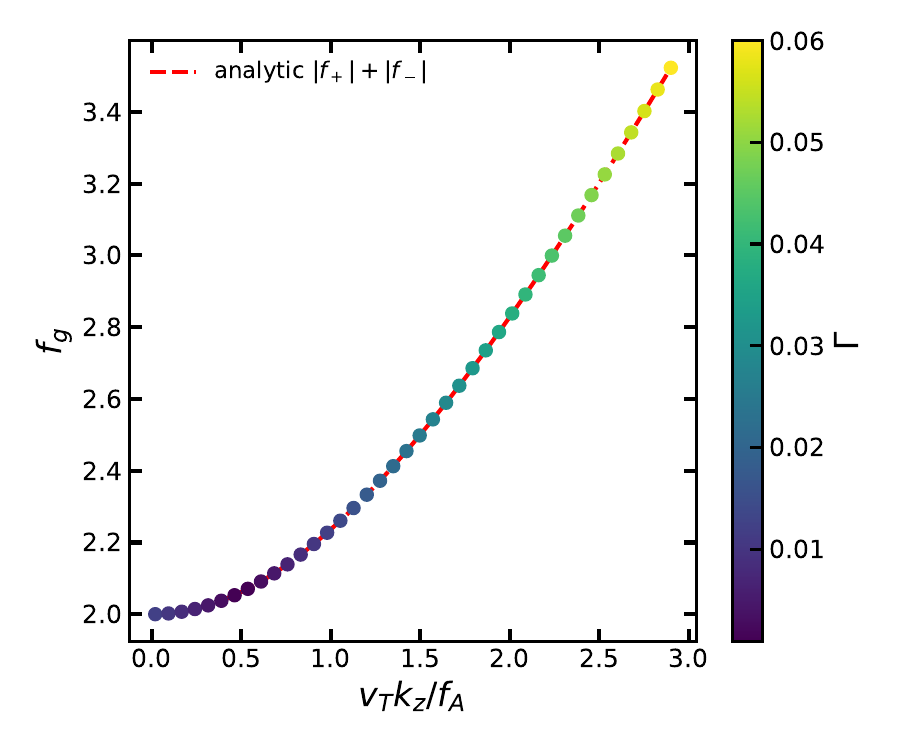}
		\caption{\raggedright Resonance location vs.\ CVE-to-Alfv\'en ratio $v_Tk_z/f_A$ (color: growth rate), with the analytic prediction $|f_+|+|f_-|$ overlaid (dashed).}
		\label{fig:cve-stochastic}
	\end{figure}
	At a fixed amplitude $h_0$, varying the GW frequency $f_g$ exposes a distinct instability band centered exactly at the sum-frequency condition, $f_g = \vert{}f_+\vert{} + \vert{}f_-\vert{}$ (Fig.~\ref{fig:3a}), where the CVE-split chiral-Alfv\'en branches serve as the coupled modes. Expanding the scan across both $f_g$ and $h_0$ gives the hallmark Arnold-tongue profile of a parametric instability, with the resonance band vanishing to a singular line as $h_0 \to 0$ and expanding monotonically with $h_0$ (Fig.~\ref{fig:3b}).
    
    Evaluating the resonance location against the CVE-to-Alfv\'en strength ratio, $v_Tk_z/f_A$ (Fig.~\ref{fig:cve-stochastic}), confirms that our numerical results match the analytical sum-frequency prediction over the full parameter range. We chose $v_Tk_z/f_A \sim \mathcal{O}(1)$ for Figs.~\ref{fig:1a} and \ref{fig:resonance-scan} to clearly resolve the branch splitting and resonance structure. Typically, this ratio is
    \begin{equation}
        \frac{v_T k_z}{f_A} \sim \frac{(\xi_0/T^2)}{\sqrt{g_{*s}}}
        \label{eq:vt-fa-ratio}
    \end{equation}
    where $g_{*s}$ is the relativistic degrees of freedom contributing to the entropy density. For $\xi_0/T^2\sim \mathcal O(1)$ and $g_{*s} \sim 100$, the ratio is of the order $10^{-3}$, and the CVE term is a small correction to ordinary Alfv\'en-wave inertia rather than comparable to it.  For the parameters $f_{\text{A}} = 1$, $v_{T} k_z = 0.3$, and $h_0 = 0.05$ (in units where $k_z = 1$), numerical Floquet analysis gives a peak growth rate of $0.005002$ at $f_g = 2.0224$. Using our full analytical formula, we obtain
    \begin{align*}
       \Gamma &= \frac{1}{2} h_0 f_{\text{A}} \left\vert{} 1 - \frac{v_{\text{A}}^2}{2} - v_{\text{T}} \right\vert{} = \frac{1}{2}(0.05)(1)\vert{}0.2\vert{} = 0.005
    \end{align*}
    agreeing with the numerical value. On the other hand, neglecting the current density terms $\mathbf{j}_{\text{E}}$ and $\mathbf{j}_{\text{B}}$ along with the GW-vorticity interaction gives an uncorrected estimate of $\Gamma_0 = \tfrac{1}{2} h_0 f_{\text{A}} = 0.025$. In the non-chiral limit ($v_{\text{T}} \to 0$), this expression simplifies to $\Gamma \to \tfrac{1}{2} h_0 f_{\text{A}} (1 - v_{\text{A}}^2 / 2) = 0.0125$, recovering the asymptotic growth rate.
	\subsection*{Chirality dependence of the growth rate}
	\label{sec-chirality-dependence}
\begin{figure*}[t]
        \centering
        \subfloat[]{\label{fig:5a}
            \includegraphics[width=0.4\textwidth,height=0.32\textwidth]{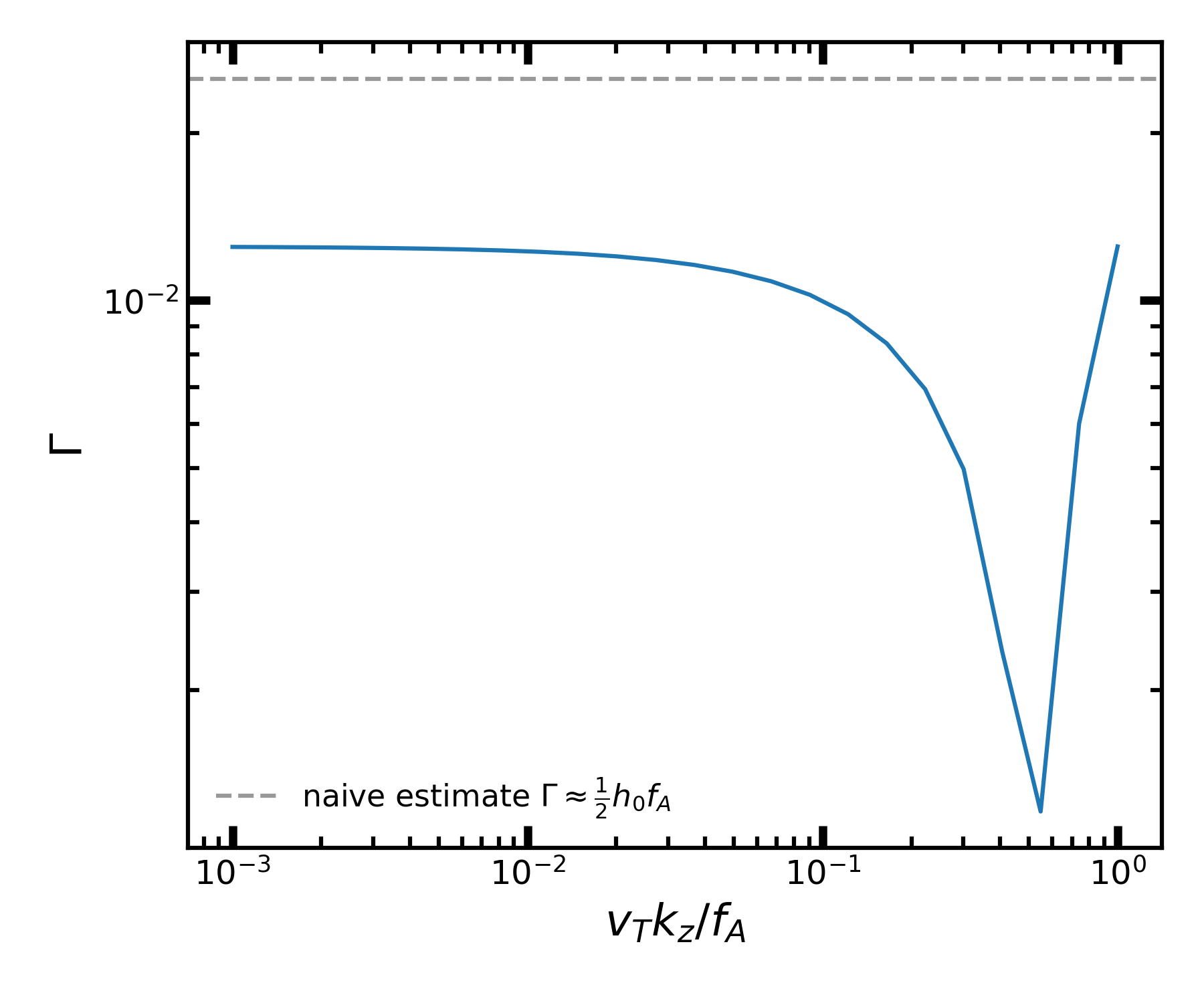}
        }
        \subfloat[]{\label{fig:5b}
            \includegraphics[width=0.42\textwidth,height=0.32\textwidth]{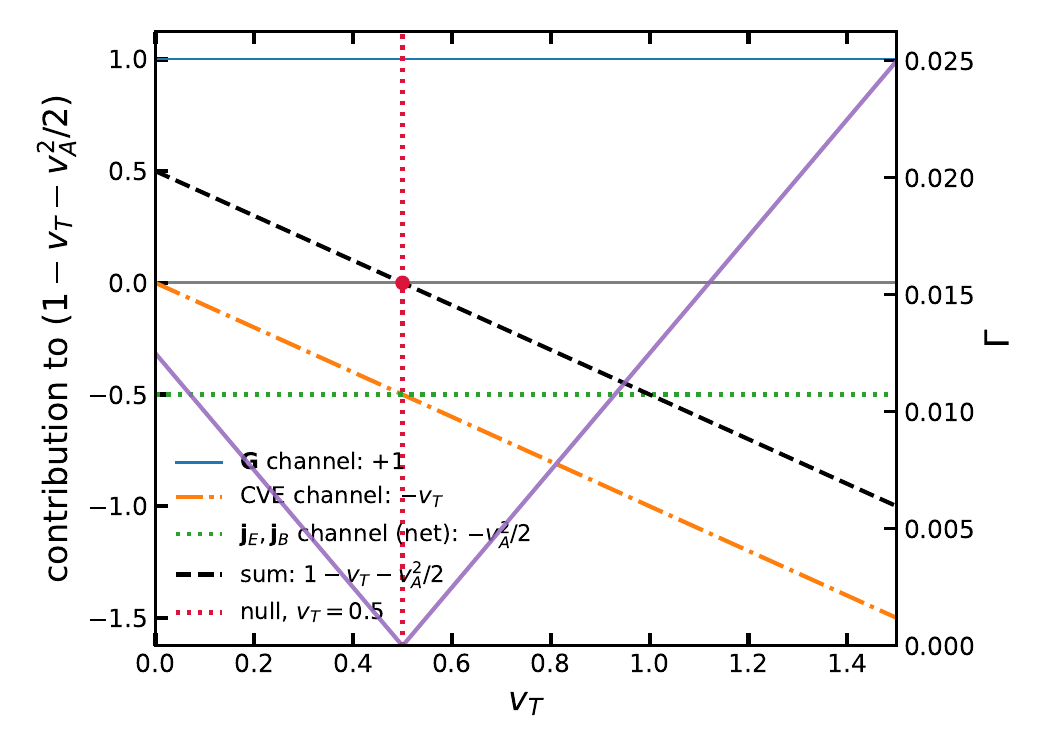}
        }
    
        \caption{\raggedright (a) Peak Floquet growth rate vs. the chirality strength $v_Tk_z/f_A$, at fixed $f_A=1,v_A=1,h_0=0.05$, against the naive dimensional estimate $\Gamma_0=\tfrac12h_0f_A$ (dashed) obtained when ${\bf j}_E,{\bf j}_B$ and the GW-vorticity term are neglected. (b) Channel decomposition of the closed-form growth rate argument $(1-v_T-v_A^2/2)$ evaluated at $v_A=1$. The total argument is broken down into a positive contribution from $\bm{G}$, a $-v_T$ term from the GW-driven vorticity/CVE channel, and a $-v_A^2/2$ net contribution from $\bm{j}_E, \bm{j}_B$. The combined sum (dashed black) crosses zero at $v_T=0.5$, completely suppressing the full growth rate $\Gamma$ (right axis). This illustrates an exact destructive interference between the active physical channels.}
        \label{fig:chirality-knob}
    \end{figure*}

    As shown in Eq.~\eqref{eq:vt-fa-ratio}, the ratio $(v_T k_z)/f_A$ is determined by the chirality strength and temperature. Plotting the peak growth rate $\Gamma$ against this ratio (Fig.~\ref{fig:5a}) reveals a strongly non-monotonic trend. $\Gamma$ begins at 0.01248 for $v_Tk_z/f_A \to 0$, plunges to a near-zero minimum near $v_Tk_z/f_A \approx 0.5$, and recovers to 0.01250 at $v_Tk_z/f_A=1$. This behavior is a consequence of the closed-form result $\Gamma=\tfrac12h_0f_A|1-v_A^2/2-v_T|$, which vanishes exactly at $v_T=1-v_A^2/2$. 
    This zero-crossing reflects the exact cancellation of three distinct physical mechanisms driving the $a_I \leftrightarrow a_{II}$ coupling in Eq.~\eqref{eq:corrected-master}, namely, (i)  a bare, plasma parameter independent positive contribution from ${\bf G}$, (ii)  $-v_T$ from the GW-driven vorticity that feeds the CVE current, and (iii) a $v_A$-dependent part from ${\bf j}_E,{\bf j}_B$. At $v_T = 1 - v_A^2/2$, the parametric drive is entirely nullified even though all three constituent channels remain active. Figure~\ref{fig:5b} illustrates this cancellation explicitly by plotting the three individual channel contributions alongside their net sum against $v_T$ at fixed $v_A=1$.  The total sum crosses zero precisely at $v_T = 0.5$, matching the sharp dip in $\Gamma$. 

\begin{figure}
		\centering
		\includegraphics[width=0.43\textwidth]{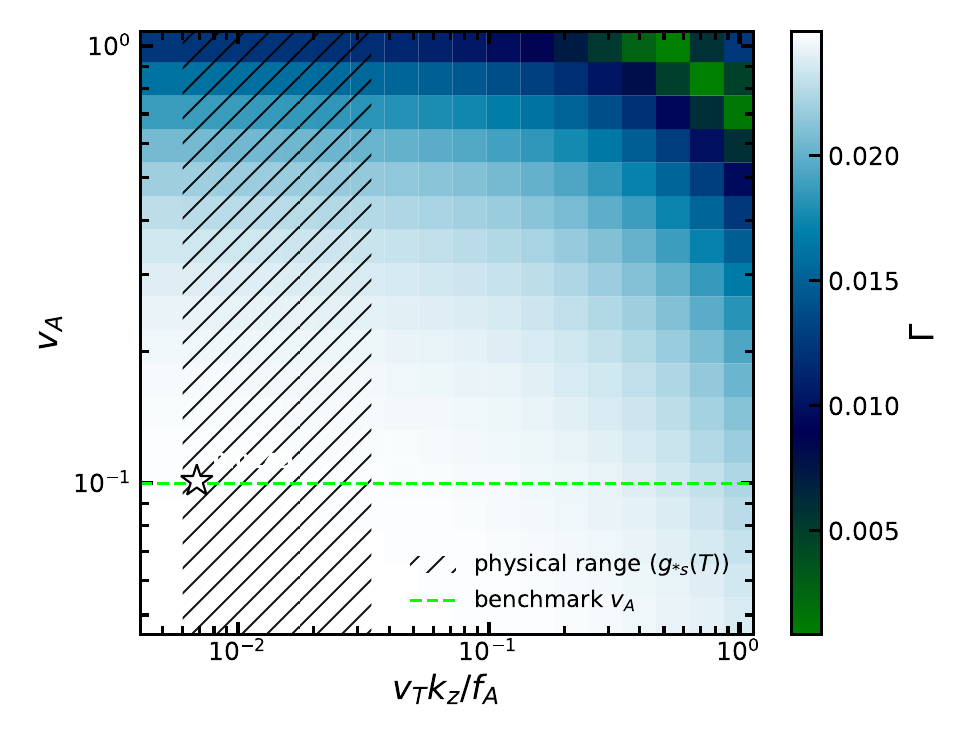}
		\caption{\raggedright The growth rate over the full $(v_Tk_z/f_A,v_A)$ plane. The hatched band marks the range $0.0061$--$0.0341$ to which $g_{*s}(T)$ pins the chirality strength across the entire radiation era. The dashed line marks the physically anchored benchmark $v_A=0.1$, and the star marks where that benchmark crosses the physical $v_Tk_z/f_A$ range.}
		\label{fig:6}
	\end{figure}

	Furthermore, $\Gamma$ is not governed only by the ratio of the chiral to Alfv\'enic terms but also exhibits dependence on both $v_Tk_z/f_A$ and $v_A$. To capture this, we map the full two-parameter space $(v_Tk_z/f_A, v_A)$ in Fig.~\ref{fig:6}, which illustrates the behavior of the growth rate. For instance, at a fixed chirality strength of $v_Tk_z/f_A \approx 6\times10^{-3}$, reducing $v_A$ from 1 to 0.1 increases the growth rate from 0.01248 to 0.02486, nearing $\Gamma/H_* = 0.025$.
    Contrary to expectation from the $v_A = 1$ regime, a smaller $v_A$ causes less suppression of the instability.
    This independent two-parameter structure is driven by the $v_A$-dependent magnetic tension and ${\bf j}_E, {\bf j}_B$ terms in Eq.~\eqref{eq:corrected-master}. This behavior plays a central role in our physically anchored standard. This can be made concrete with an order-of-magnitude estimate
    of $v_T/v_A=\xi_0/\sqrt{\epsilon_0+p_0}$. To leading order $\xi_0\sim(D/2)T^2$, and in the radiation-dominated era $\sqrt{\epsilon_0+p_0}=T^2\sqrt{(2\pi^2/45)g_{*s}}\sim7T^2$ for $g_{*s}\sim100$, giving $v_T/v_A\sim10^{-3}$. Given the analytical growth rate scaling $\Gamma \propto \vert{}1 - v_A^2/2 - v_T\vert{}$, it directly follows that the chiral $v_T$ term strictly dominates the magnetic $v_A^2/2$ term whenever $v_A \lesssim 2\times10^{-3}$.
	To determine where a physical plasma lies along this chirality axis, consider the temperature dependence of the governing parameters. It is evident from Eq.\eqref{eq:vt-fa-ratio} that the ratio $v_T / v_A $ does not run with temperature directly. Its variation is governed entirely by the effective number of relativistic degrees of freedom, $g_{*s}(T)$. Throughout the radiation-dominated era, $g_{*s}(T)$ varies by only a factor of $\sim 30$, ranging from $g_{*s} = 106.75$ in the full Standard Model regime down to $g_{*s} \approx 3.4$ following $e^+ e^-$ annihilation. This restricts $v_T k_z / f_A$ to a narrow physical range between $0.0061$ and $0.0341$ (Fig.~\ref{fig:6}). This may change provided $\xi_0/T^2$ also varies with the temperature. Sitting at $\mathcal{O}(10^{-3}\text{--}10^{-2})$, this non-zero ratio is the direct numerical manifestation of the chiral-vortical effect, defining the physical chirality threshold to which our proposed resonance mechanism is sensitive. 
    \begin{figure*}[t]
        \centering
        \subfloat[]{\label{fig:7a}
            \includegraphics[width=0.4\textwidth]{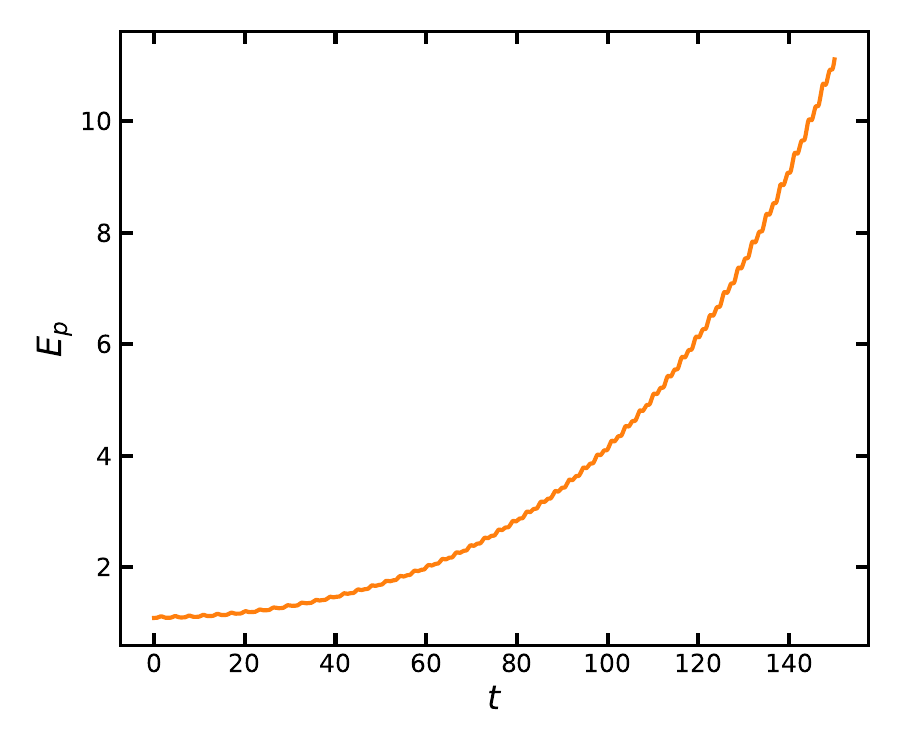}
        }
        \subfloat[]{\label{fig:7b}
            \includegraphics[width=0.43\textwidth,height=0.34\textwidth]{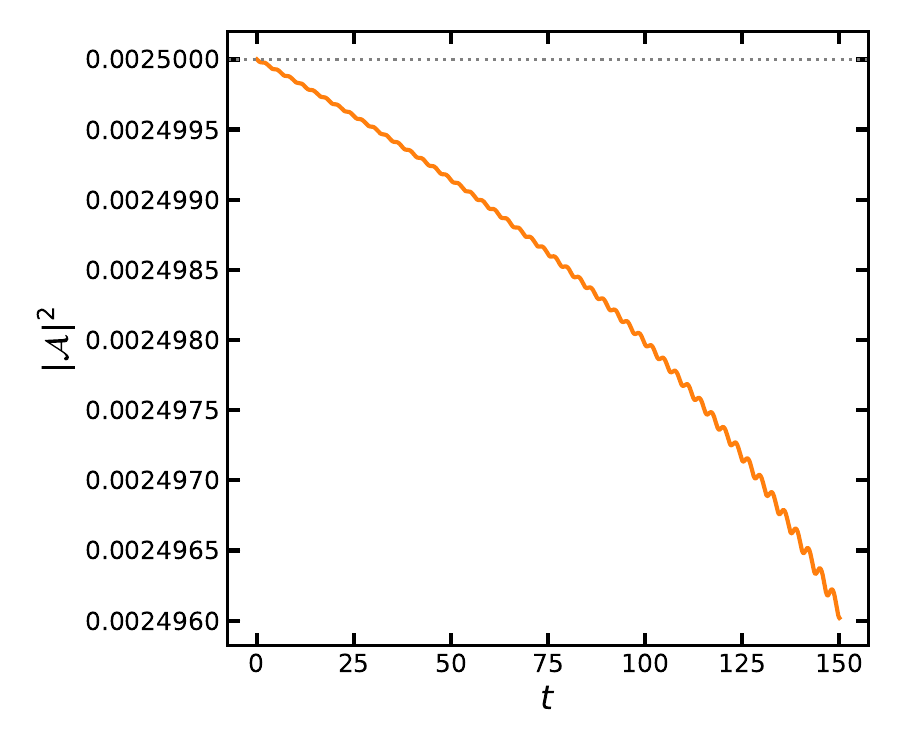}
        }
    
        \caption{\raggedright Closed backreacting system at $\chi=10^{-6}$, $a_I(0)=1$, $a_{II}(0)=0.3$, $\mathcal A(0)=0.05$. (a) Plasma energy $E_p(t)=|a_I|^2+|d_I|^2+|a_{II}|^2+|d_{II}|^2$, growing at essentially the undepleted linear rate. (b) Pump amplitude squared $|\mathcal A(t)|^2$: near-frozen until $E_p$ grows large enough to matter, then genuine depletion (dotted line: initial value).}
    		\label{fig:pump-depletion}
    \end{figure*}
	\subsection{Backreaction dynamics: depletion and finite-time blow-up}
	\label{sec-longtime}
	This subsection characterizes the onset of pump depletion and the functional dependence of the finite-time blow-up on the coupling parameter. Consistency of the closed-system dynamics incorporating backreaction was validated through several independent tests. 
    
    In the $\chi \to  0$ (undepleted-pump) limit, the closed system recovers Eq.~\eqref{eq:corrected-master} to very high precision. For the initial conditions, $a_I(0) = 1$, $a_{II}(0) = 0.3$, $d_I(0) = d_{II}(0) = 0$, $\mathcal{A}(0) = 0.05$, used in this section, the analytical initial rate of change $\left.\frac{d\vert{}\mathcal{A}\vert{}^2}{dt}\right\vert{}_{t=0}$ is negative, confirming immediate pump depletion. As shown in Fig.~\ref{fig:pump-depletion}, the full numerical trajectory exhibits clear, oscillatory pump depletion at $\chi = 10^{-6}$. $\vert{}\mathcal{A}\vert{}^2$ remains near its initial value of $0.00250$ until the plasma energy density $E_p$ grows sufficiently, eventually decreasing to $0.002496$ at $t = 150$ units as $E_p$ increases from $1.09$ to $11.11$. Initializing the system with small, randomly phased daughter seeds at a weaker coupling ($\chi = 10^{-8}$), the physically relevant regime for noise-driven instabilities yields an identical qualitative behavior across five independent random-phase realizations. Specifically, $\vert{}\mathcal{A}\vert{}^2$ remains effectively constant to several significant figures, even as $E_p$ grows by approximately three orders of magnitude in each realization. While backreaction remains present, the coupling strength is too small to produce a resolvable net trend in $\vert{}\mathcal{A}\vert{}^2$ prior to $E_p$ growing comparable to the pump reservoir.
    \begin{figure*}[t]
        \centering
        \subfloat[]{\label{fig:8a}
            \includegraphics[width=0.43\textwidth]{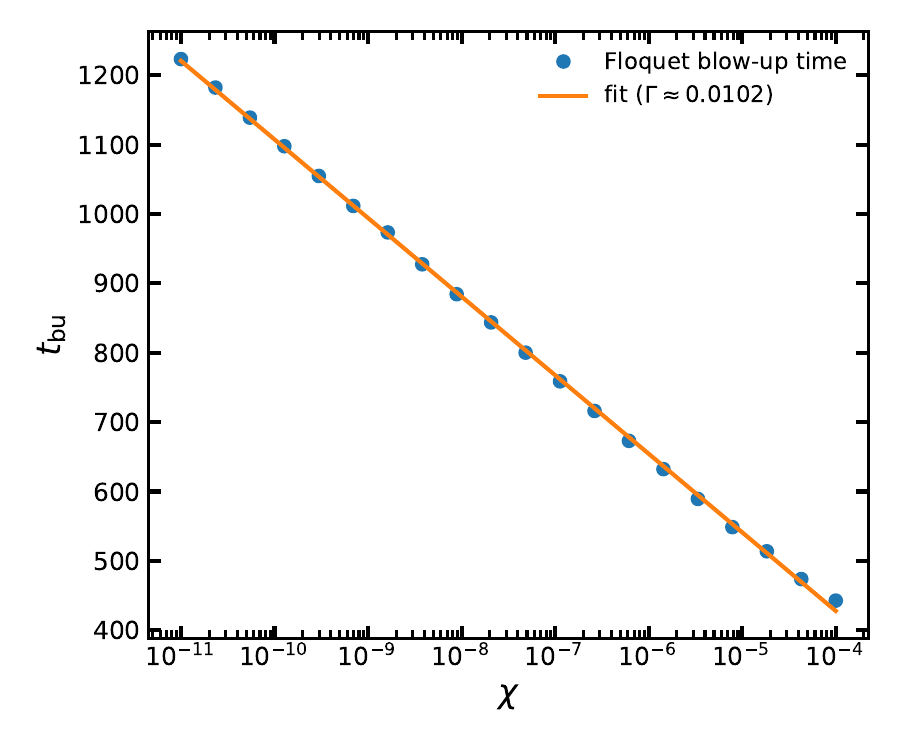}
        }
        \subfloat[]{\label{fig:8b}
            \includegraphics[width=0.43\textwidth]{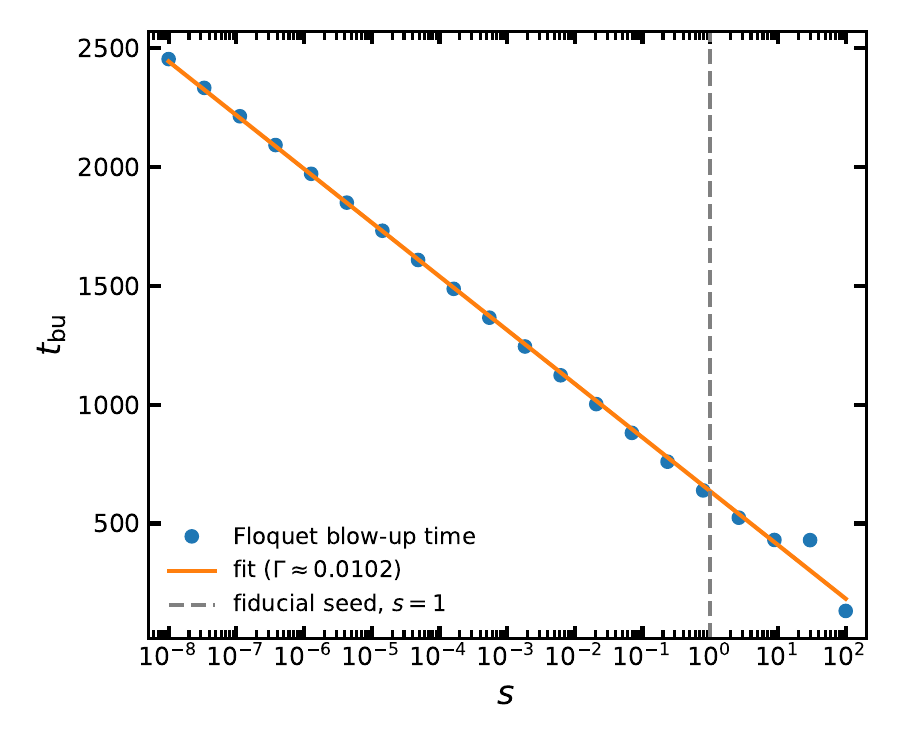}
        }
    
        \caption{\raggedright Blow-up time for the closed system. The panels show two independent 1D projections of the same logarithmic scaling law. (a) Dependence on $\chi$ over seven decades at a fixed seed. The empirical fit, $t_{\rm bu}\approx-49.2\ln\chi-25.0$, corresponds to a growth rate of $\Gamma\approx0.0102$. (b) Dependence on the daughter-mode seed rescaling $s$ ($a_I(0)=s, a_{II}(0)=0.3s$) over ten decades at a fixed $\chi=2\times10^{-6}$. The fit $t_{\rm bu}\approx-98.2\ln s+635.3$ results in the exact same $\Gamma\approx0.0102$, confirming that both slices probe the same underlying law.
        }
    		\label{fig:longtime-blowup}
    \end{figure*}
	Consistent with this, a simple analytic invariant of the form $E_p+\kappa|\mathcal A|^2={\rm const}$ does not hold with fixed $\kappa$ once backreaction does become significant. The local slope $\kappa=-dE_p/d|\mathcal A|^2$, evaluated at four points along a $\chi=2\times10^{-6}$ trajectory, takes values differing by more than two orders of magnitude and in sign ($4.3\times10^3$, $2.0\times10^5$, $1.7\times10^6$, $-2.7\times10^5$). Both findings trace to the same cause that $a_I, a_{II}$ each carry both natural frequencies $f_\pm$ simultaneously, so neither the single-frequency Manley--Rowe relation \cite{manley1956some} nor a simple monotonic depletion picture directly transfers.

	A comprehensive suite of robustness tests was previously performed for a simplified two-variable model. These tests included a multi-decade parameter scan in $\chi$, spectral purity assessments of the Manley-Rowe breakdown, and ensemble statistics over random initial phases. We did not fully replicate these exhaustive checks for the four-variable system presented here. Nevertheless, because the backreaction source term remains mathematically identical, the qualitative dynamics are expected to persist. Specifically, we anticipate physical depletion in the weak-coupling regime alongside the absence of a simple conserved quantity. A full quantitative replication of these robustness tests for the four-variable system is deferred to future work.
    
	Contrary to the intuitive expectation that weaker coupling merely stabilizes the system, the numerical trajectories presented above--all evaluated up to $t_{\max} \lesssim 400$-- do not represent the ultimate long-term fate of the coupled system. Extending the integration range reveals a fundamental feature. For the specified initial conditions ($a_I(0)=1, a_{II}(0)=0.3, d_I(0)=d_{II}(0)=0, \mathcal{A}(0)=0.05$), Fig.~\ref{fig:pump-depletion} reveals that there exists no non-zero coupling threshold below which the closed system remains bounded. For every tested coupling $\chi \in [10^{-11}, 10^{-4}]$, the system eventually exhibits a finite-time blow-up, with the blow-up time $t_{\text{bu}}$ diverging only logarithmically as $\chi \to 0$. At the singularity, the integrator step size collapses to zero as the field amplitudes $\vert{}a_I\vert{}$, $\vert{}a_{II}\vert{}$, and $\vert{}\mathcal{A}\vert{}$ diverge simultaneously within finite $t$. Only the exact uncoupled limit ($\chi = 0$) remains defined indefinitely, reducing to the standard linear regime with exponential growth but no finite-time divergence. Critically, whether this instability occurs on a timescale of a few Hubble times or several thousand depends on the ratio of $\chi$ to the competing growth rate $\Gamma$. This proves essential when evaluating physical cosmological parameters.

	The blow-up time $t_{\rm bu}(\chi)$ follows a clean logarithmic law over the range tested (Fig.~\ref{fig:8a}),
	\(
	   t_{\rm bu}(\chi) \approx -\frac{1}{2\Gamma}\ln\chi + {\rm const}\, ,
	\)
	with the fitted slope corresponding to $\Gamma\approx0.0102$. This relationship emerges directly from first principles. Since the mode product $a_I\overline{a_{II}}$ grows exponentially as $e^{2\Gamma t}$ irrespective of how small $\chi$ is, the timescale required for the non-linear coupling term $\chi a_I\overline{a_{II}}$ to reach $\mathcal{O}(1)$ is determined entirely by the growth rate $\Gamma$.  Our numerically extracted $\Gamma \approx 0.0102$ matches the analytical limit $\Gamma = \frac{1}{2}h_0 f_A \vert{}1 - v_A^2/2 - v_T\vert{} = 0.01$ evaluated at the effective drive amplitude $h_\times = 2\mathcal{A}(0) = 0.1$.
    By symmetry, this mechanism extends to the daughter-mode seed amplitude. Since the initial amplitudes factor into the exponential evolution as $a_I\overline{a_{II}} \propto \vert{}a_I(0)a_{II}(0)\vert{} e^{2\Gamma t}$, rescaling the initial seed ($a_I(0)=s, a_{II}(0)=0.3s$) shifts the blow-up time by $\Delta t_{\rm bu} = -\Gamma^{-1} \ln s$ at constant $\chi$. This scaling relation is verified across ten orders of magnitude in $s$ (Fig.~\ref{fig:8b}), resulting in an independent extraction of $\Gamma \approx 0.0102$. The exact agreement between these distinct parameter scans confirms that both numerical results reflect a single underlying dynamical law.

	Instead of reaching a bounded asymptotic state, the system exhibits a finite lifetime that scales only logarithmically as $\chi \to 0$. This runaway growth eventually breaks the small-perturbation assumption. Therefore, rather than implying a physical divergence, this behavior shows that the quadratically truncated equations lack an intrinsic saturation mechanism. Identifying the true asymptotic state requires either incorporating higher-order nonlinearities or introducing external physical cutoffs (such as wave-breaking or turbulent dissipation) outside our current framework. Consequently, this finding reshapes our approach to the subsequent physical analysis. The condition for backreaction to remain negligible over a given timeframe is not that $\chi$ is small in absolute terms, but that it is small compared to the growth rate $\Gamma$. Importantly, this suppression does not trivially scale as $v_A \to 0$, contrary to expectations derived from the $v_A = 1$ regime.
	\begin{figure}[!t]
		\centering
		\includegraphics[width=0.45\textwidth]{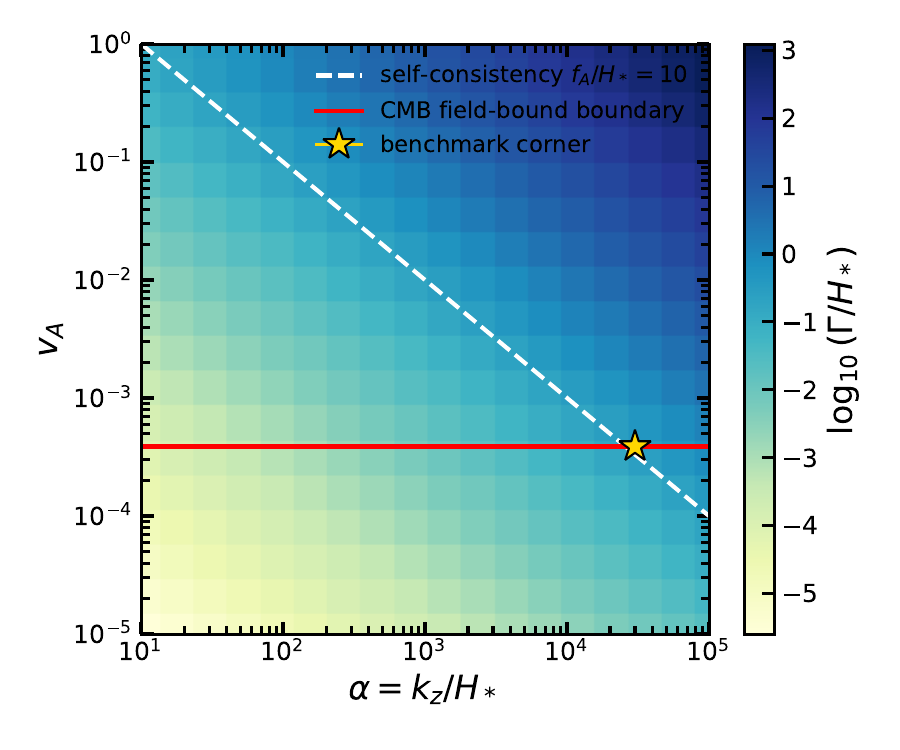}
		\caption{\raggedright Floquet growth rate over the $(\alpha,v_A)$ plane, with the self-consistency boundary $f_A/H_*=10$ (white) and the CMB field-bound boundary (red) overlaid. The two constraints cross at $\alpha\approx2.6\times10^4$. The allowed region is the wedge above the white line and below the red line, extending indefinitely toward larger $\alpha$ and smaller $v_A$. The gold star marks the selected parameter point.}
		\label{fig:alpha-vA-map}
	\end{figure}
	%
	\subsection{Physical analysis and self-consistency}
	\label{sec-results-benchmark}
	Mapping the dimensionless parameters onto an early-universe plasma requires $f_A, f_g \gg H_*$ to validate the fixed-background approximation. Additionally, the implied field strength must satisfy existing nanogauss-level CMB bounds on primordial magnetic fields~\cite{planck2016planck}. Both constraints depend solely on the two ratios $\alpha = k_z/H_*$ and $v_A$. Specifically, the field strength bound relies solely on $v_A$, as the $T_*$-dependence cancels precisely between production and redshifting. The self-consistency reduces to the combination $f_A/H_* = v_A\alpha$. Consequently, both conditions can be evaluated across the full $(\alpha, v_A)$ parameter plane. We adopt $f_A/H_* = 10$ as the operational boundary representing $f_A/H_* \gg 1$. Rather than an arbitrary threshold, this value ensures that the backreaction coupling is suppressed to a few percent of the Hubble rate, $\chi/H_* = 1 / [4(f_A/H_*)] \approx 0.025$, thereby providing a quantitative criterion for the self-consistency of the fixed-background approximation.
	
	An initial horizon-scale choice ($\alpha=1$) strictly violates the self-consistency requirement ($f_A/H_*=0.1$). Moving to a sub-horizon scale with $\alpha=100$ and $v_A=0.1$ reaches the self-consistency threshold ($f_A/H_*=10$), though without a significant margin. Moreover, it exceeds the present-day CMB magnetic field bound by two orders of magnitude ($B_0\sim2\times10^{-7}$~G). Adjusting $v_A$ downward to satisfy the CMB limit while holding $\alpha=100$ fixed, motivated by the scaling $f_0\propto v_A$ at fixed $(\alpha, T_*)$, does not resolve the tension. It reduces $f_A/H_*$ by the same factor to $0.039$, merely trading the field-strength violation for a breakdown in self-consistency.
    \begin{figure}[!t]
    	\centering
    	\includegraphics[width=0.4\textwidth, height=0.345\textwidth]{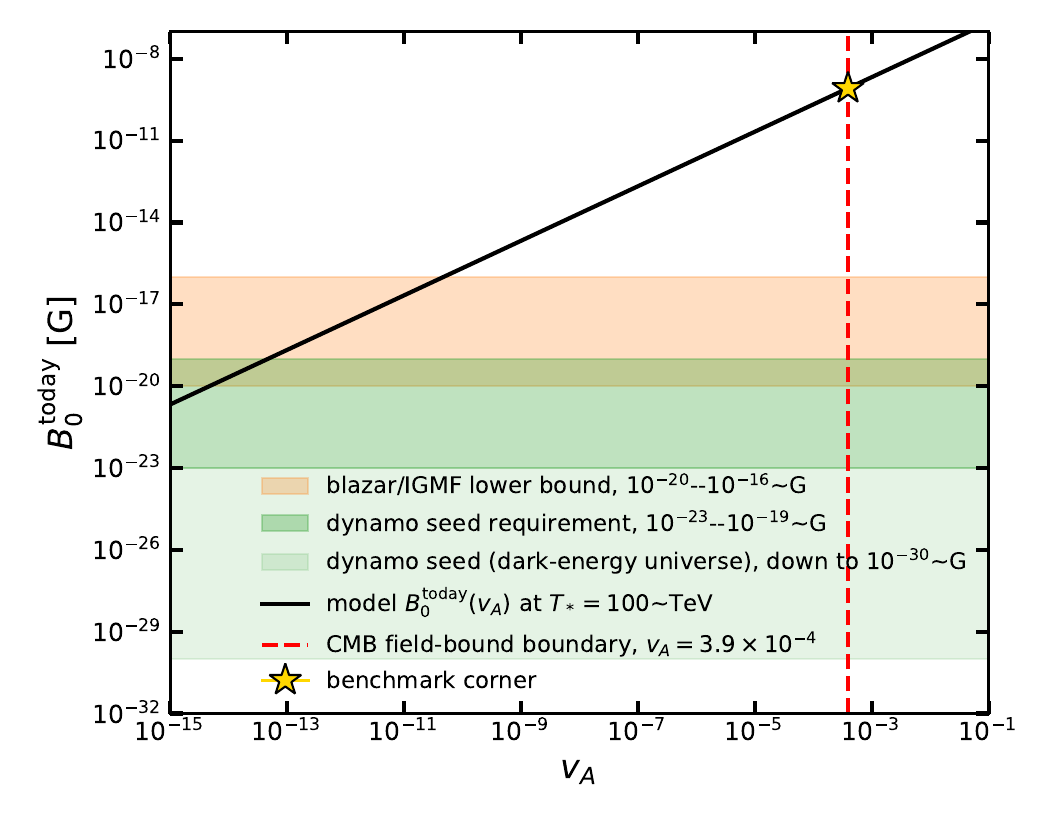}
    	\caption{\raggedright The model's today-redshifted field $B_0^{\rm today}$ vs.\ $v_A$ at fixed $T_*=100$~TeV (black line; linear in $v_A$ at fixed $T_*,g_{*s}$), against the blazar/IGMF lower bound~\cite{neronov2010evidence} (orange) and the galactic-dynamo seed-field requirement~\cite{davis1999relaxing} (green, with the dark-energy-relaxed extension shown lighter). The gold star indicates the same reference point.}
    	\label{fig:blazar-bound}
    \end{figure}
	Only $\alpha\gtrsim2.6\times10^4$ admits a $v_A$ satisfying both constraints together. Rather than an isolated solution, this threshold marks the vertex of an allowed region in the $(\alpha, v_A)$ parameter plane, mapped in Fig.~\ref{fig:alpha-vA-map}. The self-consistency boundary ($f_A/H_* = 10$, white) scales as $v_A = 10/\alpha$, whereas the CMB field-strength bound (red) forms a horizontal line at $v_A \approx 3.9 \times 10^{-4}$, reflecting its independence from $\alpha$. These two contours intersect at $\alpha \approx 2.6 \times 10^4$. Beyond this crossing point, the permitted region forms an open wedge extending toward larger values of $\alpha$, with no upper limit on $\alpha$ established within the scope of the present analysis.
	
	We select $\alpha = 3 \times 10^4$ and $v_A = 3.9 \times 10^{-4}$ (at $T_* = 100\text{ TeV}$, $g_{*s} = 106.75$) as our reference point because it represents the minimal deviation from the initial $(\alpha = 100, v_A = 0.1)$ values required to jointly satisfy both constraints. This reference point gives $f_A/H_* \sim 12$, ensuring self-consistency and a present-day field strength of $B_0 \sim 8 \times 10^{-10}\text{ G}$, complying with CMB limits. Redshifted using standard radiation-era expressions~\cite{caprini2020detecting}, the corresponding resonant frequency is $f_0 \sim 6 \times 10^{-2}\text{ Hz}$. This result directly connects back to the magnetogenesis motivation. Determined by the CMB upper bound rather than deliberate tuning, this comoving field strength comfortably exceeds both the lower limits on intergalactic fields derived from blazar non-observations ($\sim 10^{-20}-10^{-16}\text{ G}$ based on the absence of a GeV-band Fermi cascade~\cite{neronov2010evidence}) and the seed fields required for galactic dynamos ($\sim 10^{-23}-10^{-19}\text{ G}$, relaxing to $\sim 10^{-30}\text{ G}$ under dark energy models~\cite{davis1999relaxing}) by roughly seven to ten orders of magnitude (Fig.~\ref{fig:blazar-bound}). Consequently, this reference point constitutes more than a consistency exercise. It identifies a physically viable parameter space relevant to the origin of cosmic magnetic fields~\cite{anand2017chiral}. Notably, the peak frequency remains nearly identical to that of the unconstrained $\alpha = 100$ case because $f_0 \propto v_A \alpha$, and both parameter pairs share similar products ($v_A\alpha = 10$ versus $12$), but now without violating observational or theoretical bounds.
	
	At this reference point, the backreaction coupling remains small relative to the fast oscillation frequency $f_A$. This result follows analytically. Since $v_T k_z / f_A \ll 1$ in this regime, the resonance occurs at $f_g = \vert{}f_+\vert{} + \vert{}f_-\vert{} \approx 2f_A$, leading to 
    \begin{align}
        \frac{\chi}{f_A} &= \frac{H_*^2}{2f_gf_A} \approx \frac{H_*^2}{4f_A^2} = \frac{1}{4\,(f_A/H_*)^2}\, ,
            \label{eq:chi-over-fA}\\
        \frac{\chi}{H_*} &= \frac{f_A}{H_*}\cdot\frac{\chi}{f_A} \approx \frac{1}{4\,(f_A/H_*)}\, .
            \label{eq:chi-over-Hstar}
    \end{align}
    For $f_A/H_* \sim 12$, Eq.~\eqref{eq:chi-over-fA} gives $\chi/f_A \sim \times 10^{-3}$.
    
	This reference point serves as a representative baseline rather than a unique prediction. Without a specific production mechanism, $\alpha$ and $T_*$ are free parameters, and $f_0$ scales linearly with both. We next examine the consequences of this scaling for the predicted frequency and its experimental prospects.
	\subsection*{Detectability and reprocessing of a resonant background}
	\label{sec-detectability}
\begin{figure}[!t]
		\centering
		\includegraphics[width=0.45\textwidth,height=0.37\textwidth]{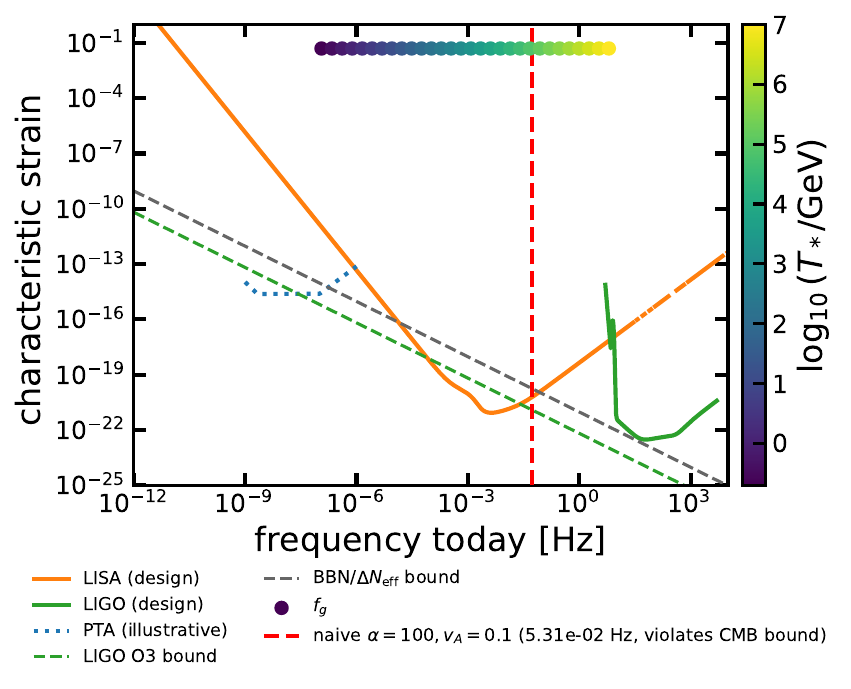}
		\caption{\raggedright Resonant frequency and detector sensitivities, colored by production temperature $T_*$ at the doubly allowed corner ($\alpha=3\times10^4, v_A=3.9\times10^{-4}$). Point heights indicate the Floquet pump amplitude ($h_0=0.05$), rather than a predicted strain. The red dashed line marks the naive, physically excluded $(\alpha, v_A)=(100, 0.1)$ reference case. Its proximity to the doubly allowed points shows that enforcing the CMB and self-consistency constraints costs essentially nothing in detectability.}
		\label{fig:multi-era-stress}
	\end{figure}

    We examine the temperature $T_*$ over the range $[0.2~{\rm GeV}, 10^7~{\rm GeV}]$, extending from near the QCD scale up to a representative Beyond-the-Standard-Model (BSM) scale. We also fix the parameters at $\alpha=3\times10^4$, $v_A=3.9\times10^{-4}$. This variation shifts the observed frequency from $1.1\times10^{-7}$~Hz to $6.2$~Hz (see colored points in Fig.~\ref{fig:multi-era-stress} and Table~\ref{tab:eras-summary}). Crucially, the model remains self-consistent and satisfies CMB bounds across this entire range. This broad frequency span traverses several primary observational windows. It fully encompasses the LISA band, overlaps with Advanced LIGO, and reaches the sensitive frequency regimes targeted by next-generation ground-based detectors like the Einstein Telescope and Cosmic Explorer~\cite{punturo2010einstein,reitze2019cosmic}. 
    Moving toward larger $\alpha$ and smaller $v_A$ would drive the observed frequency even higher. Given the scaling relation $f_0 \propto v_A\alpha$, this product strictly increases when moving away from the corner of the allowed parameter space in Fig.~\ref{fig:alpha-vA-map}. Consequently, the frequency is not a sharp prediction even at a fixed $T_*$. While the corner of the allowed region establishes a firm lower bound, this work does not establish an upper bound.

\begin{figure}
        \centering
        \includegraphics[width=0.45\textwidth]{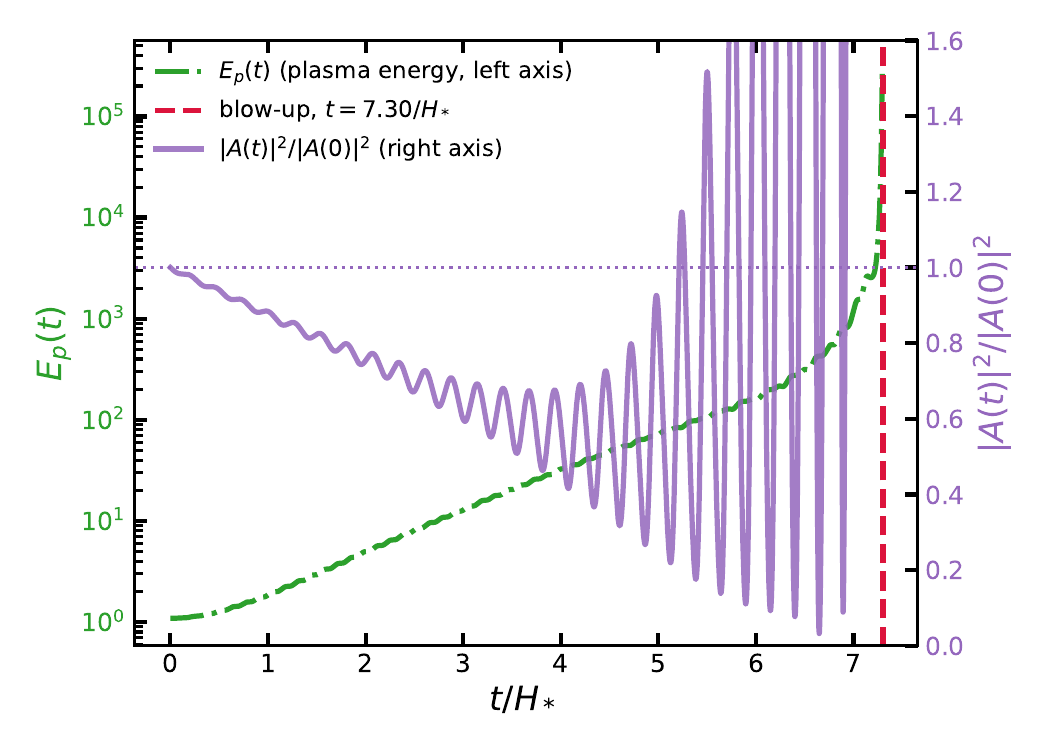}
        \caption{\raggedright Plasma energy $E_p(t)$ (green, left axis, log) and fractional squared pump amplitude $\vert{}\mathcal A(t)\vert{}^2/\vert{}\mathcal A(0)\vert{}^2$ (purple, right axis) evaluated at physical coupling $\chi=H_*/(2f_g)$ for $T_*=10^5$~GeV. A finite-time blow-up occurs at $t\approx7.3/H_*$, consistent across all tested epochs. The numerical integration is plotted up to the divergence (dashed line).}
        \label{fig:energy-exchange-eras}
    \end{figure}

	Figure~\ref{fig:multi-era-stress} indicates the spectral location of the resonance, not its observable amplitude. Predicting the absolute strain requires specifying a primordial GW production mechanism to seed the resonance, which is beyond the scope of this work. However, the backreaction analysis allows us to formulate a quantitative constraint. As demonstrated below, the closed system enters the nonlinear regime,  characterized by order-unity fractional energy exchange, within $t_{\rm bu} \sim 7$ to $90/H_*$ for all tested epochs, depending on the initial seed. This timescale is well under one hundred Hubble times, contradicting the assumption that backreaction acts only as a small correction to linear growth over many $e$-folds. Direct integration of the closed system using the physical coupling for this selected parameter set confirms this behavior. Specifically, the pump amplitude $\vert{}\mathcal A(t)\vert{}^2$ decreases to $3.3\%$ of its initial value at $t \approx 6.65/H_*$, shortly before the truncated equations diverge at $t_{\rm bu} \approx 7.30/H_*$ (see Fig.~\ref{fig:energy-exchange-eras}). Therefore, any background present at the resonant frequency undergoes order-unity reprocessing on this timescale. The only undetermined factor is the overall normalization at $f_0$, which is dictated by the background source rather than the resonance mechanism. Consequently, while Fig.~\ref{fig:multi-era-stress} does not predict an absolute strain, it identifies the frequency band where a primordial signal, if present, would undergo significant modification.

	This consistency extends across the entire radiation era. Since the chirality strength $v_Tk_z/f_A$ and the self-consistency ratio $f_A/H_*=v_A\alpha$ depend on the production epoch solely through $g_{*s}(T)$, the core dynamics are largely temperature-independent. Scanning $T_*\in[0.2,10^7]$~GeV at the doubly allowed corner confirms this: $f_A/H_*$ remains $11.70$, $\Gamma/H_*$ stays constant at $0.293$, and the present-day magnetic field $B_0^{\rm today}$ varies by merely $9\%$ (Table~\ref{tab:eras-summary}). Therefore, even as the plasma enthalpy $w\propto T_*^4$ scales over $31$ orders, the fundamental strength of the instability remains unchanged; the production epoch merely dictates the resonance's spectral position.

    A key consequence is that the parameters $\chi/H_*$ and $\Gamma/H_*$ are entirely fixed by $(\alpha, v_A, g_{*s}(T))$. Using our chosen parameter values, we find $\chi/H_* \approx 2.14\times10^{-2}$ and $\Gamma/H_* \approx 0.293$. Since the ratio $\chi/\Gamma \approx 0.073$ is small, backreaction constitutes a genuine but negligible correction to the unsuppressed linear growth previously derived. This also restricts the dimensionless blow-up time to a nearly era-independent value of $t_{\rm bu} \approx 7.30\text{--}7.31/H_*$ over the entire evaluated $T_*$ range. However, translating this to physical time ($t_{\rm phys} = t_{\rm bu}/H_*$) reveals an onset duration that spans roughly 15 orders of magnitude, from $1.1\times10^{-4}$~s at the QCD scale to $3.4\times10^{-20}$~s at $T_* = 10^7$~GeV (Table~\ref{tab:eras-summary}). Ultimately, the mechanism drives order-unity pump depletion in a few Hubble times at any epoch, but the physical duration of the process acts as a direct temporal signature of that epoch. This dynamic remains insensitive to the initial seed amplitude $s$. Following the relation $\Delta t_{\rm bu} = -\Gamma^{-1}\ln s$, a seed value ten orders of magnitude smaller than the fiducial conditions ($a_I(0)=1, a_{II}(0)=0.3$) shifts $t_{\rm bu}$ to only $\sim 86/H_*$, preserving the rapid completion of the exchange.

	\begin{table*}[t]
		\begin{ruledtabular}
			\begin{tabular}{lccccccc}
				Era & $T_*$~[GeV] & $g_{*s}$ & $f_0^{\rm today}$~[Hz] & $\Gamma/H_*$ & $w$~[GeV$^4$] & $t_{\rm phys}$~[s] & $B_0^{\rm today}$~[G] \\
				\hline
				near-QCD & $2.0\times10^{-1}$ & $63.1$ & $1.1\times10^{-7}$ & $0.293$ & $4.4\times10^{-2}$ & $1.1\times10^{-4}$ & $9.1\times10^{-10}$ \\
				electroweak & $1.5\times10^{2}$ & $106.7$ & $9.3\times10^{-5}$ & $0.293$ & $2.4\times10^{10}$ & $1.5\times10^{-10}$ & $8.3\times10^{-10}$ \\
				reference ($100$~TeV) & $1.0\times10^{5}$ & $106.75$ & $6.2\times10^{-2}$ & $0.293$ & $4.7\times10^{21}$ & $3.4\times10^{-16}$ & $8.3\times10^{-10}$ \\
				high-scale & $1.0\times10^{7}$ & $106.75$ & $6.2$ & $0.293$ & $4.7\times10^{29}$ & $3.4\times10^{-20}$ & $8.3\times10^{-10}$ \\
			\end{tabular}
		\end{ruledtabular}
        \caption{\raggedright Summary of epoch-dependent quantities at the allowed parameter-space ($\alpha=3\times10^4$, $v_A=3.9\times10^{-4}$), at four representative production temperatures. $f_0$ is the resonant frequency, $\Gamma/H_*$ is the growth rate, $w$ is the plasma enthalpy at production, $t_{\rm phys}=t_{\rm bu}/H_*(T_*)$ is the real-time duration of the backreaction-driven energy exchange, $B_0^{\rm today}$ is the redshifted field strength consistent with CMB at every epoch by construction.}
        \label{tab:eras-summary}
	\end{table*}
	\section{Conclusion}
	\label{sec-conclusion}
	We have derived, from the covariant equations of chiral magnetohydrodynamics coupled to linearized general relativity, a closed set of equations that describe the resonant interaction between a gravitational wave and the chiral Alfv\'en wave in a chiral plasma. The central results of this work are:
	\begin{itemize}
            \item A closed, four-variable master system (eq.~\eqref{eq:corrected-master}), derived self-consistently in the fluid's local tetrad frame. This system couples the two chiral-Alfv\'en velocity polarizations and the two magnetic-perturbation polarizations through a prescribed GW waveform, $h_\times(t)$. We also derive a simplified CVE-only limit (eqs.~\eqref{eq:cve-only-AI} and \eqref{eq:cve-only-AII}) that exactly reproduces Yamamoto's chiral Alfv\'en wave~\cite{yamamoto2015chiral} in the appropriate regime. Finally, we provide an independent derivation of the resonant growth rate using a slowly varying envelope approximation, matching the Floquet result.

		\item Numerical confirmation of a parametric instability at the sum-frequency condition $f_g=\vert{}f_+\vert{}+\vert{}f_-\vert{}$, which exhibits characteristic Arnold-tongue broadening with GW strain.
        
		\item An explicit identification of the chirality strength $v_Tk_z/f_A$ directly with the CVE transport coefficient. We demonstrate that this ratio is not a free parameter in a realistic early-Universe plasma as the temperature dependencies of $v_T$ and $v_A$ cancel exactly and the ratio is determined solely by $g_{*s}(T)$. Consequently, it is tightly constrained to the range $0.006$--$0.034$ throughout the entire radiation era. Furthermore, we show that the peak growth rate depends independently on $v_A$, revealing a two-parameter dependence.
        
		\item We developed a closed, five-variable framework to capture the nonlinear backreaction and energy exchange between the gravitational wave and the plasma. Because the resulting daughter modes are not spectrally pure, standard Manley–Rowe invariants and simple pump-depletion models are inapplicable. At our physically self-consistent parameter set, the requisite small values of $v_A$ eliminate the chirality-dependent suppression term, allowing the linear growth rate $\Gamma$ to remain near its unsuppressed maximum. Consequently, backreaction acts only as a minor perturbation. Despite this, the nonlinear system consistently diverges in finite time—within approximately seven Hubble times across all tested epochs. This behavior confirms that the system's eventual breakdown is dictated by the potent unsuppressed linear resonance rather than by backreaction terms.
        
		\item We identify a parameter regime ($\alpha=3\times10^4, v_A=3.9\times10^{-4}$) that is both sub-horizon and CMB-compliant identically across the entire radiation era. While the dimensionless growth rate remains fixed at $\Gamma/H_*=0.293$ (Table~\ref{tab:eras-summary}), changing the production epoch causes the resonant frequency to redshift by eight orders of magnitude, the plasma's stress-energy content to span thirty-one orders of magnitude, and the real-time duration of the backreaction to shift by fifteen orders of magnitude. This demonstrates that the production epoch dictates not just the resonant frequency, but the fundamental timescale of the energy exchange between the gravitational wave and the plasma.
        
	\end{itemize}
	The present mechanism belongs to a broader class of parametric GW resonances sourced by periodic cosmological backgrounds. For instance, Ref.~\cite{cai2024parametric} recently demonstrated that an oscillating scalar field in a modified-gravity (DHOST) framework can drive a GW mode into a Hill's-equation-type instability. While both mechanisms depend on the resonance-frequency-to-Hubble-rate ratio to determine in-band duration, the fundamental coupling direction is reversed. In Ref.~\cite{cai2024parametric}, the GW is the driven field and the primary observable. Here, operating in standard general relativity, an ambient GW drives the chiral plasma's daughter modes, and we explicitly track the resulting depletion of the GW pump via nonlinear backreaction. Ultimately, this comparison highlights that a narrow, sharp peak in a GW spectrum is not a unique smoking gun. Both gravity modifications and anomalous plasma transport can generate such features; distinguishing between them requires secondary evidence, such as a primordial magnetic-field signature for the latter, or a scalar-field signature for the former.
	
	Several theoretical extensions are necessary before this mechanism can provide quantitative, falsifiable predictions for current  GW or magnetic-field observations. We leave the following open challenges for future work:
    \begin{itemize}
    \item Resolving the multi-frequency backreaction closure to determine the true saturated amplitude, rather than just its sign. This must be accompanied by comprehensive robustness testing for the four-variable master system, analogous to the tests performed on the simpler two-variable model.
    \item Extending the four-variable master system (eq.~\eqref{eq:corrected-master}) to handle resonance crossing under chirping or otherwise non-periodic drives. Because the Floquet theory utilized here applies solely to strictly periodic drivers, the crossing rate must ultimately be tied to a self-consistent astrophysical or cosmological source.
    \item Investigating whether $\xi_B$ is genuinely negligible in a realistic, multi-species early Universe plasma, rather than relying on the formally subleading assumption of our single-species treatment.
    \item Anchoring the assumed GW and plasma parameters ($v_A$, $v_T$, $\alpha$, and $h_0$) to a specific, ultraviolet-complete physical mechanism—such as a first-order electroweak phase transition or a chiral-plasma-instability seed—rather than treating them as free variables.
    
    \item Relaxing the $h_+ = 0$ restriction adopted throughout this study to determine whether a combined treatment of both GW polarizations alters the resonance dynamics.
    \end{itemize}
	%
	
	\appendix
	\section{Derivation of the GW-sourced coupling channels}
	\label{app:coupling-channels}
	This appendix derives two further GW-sourced coupling channels used throughout Sec.~\ref{sec-GW-wave-inter}, beyond the force ${\bf G}$. %
	The gravitational-wave-induced terms introduced in Sec.~\ref{sec:-II} are formulated by contracting this connection with either the fluid four-velocity or the field strength tensor. Unlike a torsion-free connection evaluated in a coordinate basis, this connection is not symmetric in its lower indices. Because the tetrad basis vectors, $e_1$ and $e_2$, do not commute, their anholonomy leads to $\omega^c{}_{ab} - \omega^c{}_{ba} = C^c{}_{ab} \neq 0$. For instance, $\nabla_0 e_1 = 0$, whereas $\nabla_1 e_0 = \tfrac12 \partial_t h_+ \, e_1 + \tfrac12 \partial_t h_\times \, e_2$. While these quantities must be identical for a symmetric connection, they diverge in this non-coordinate frame. It is precisely this fundamental asymmetry that generates the two effects described below.
    \begin{itemize}
        \item[(i)]  A generalized-Maxwell source, ${\bf j}_E$: The frame-dependent term ${\bf j}_E$, given in Eq.~\eqref{eq:j_e}, relies on the derivatives $\dot h_+$ and $\dot h_\times$ (taken with respect to $\xi=z-t$) and couples to the perturbed fields $\delta{\bf E}$ and $\delta{\bf B}$ rather than the background state. This establishes a distinct, physical coupling channel of the same perturbative order, $O(h\,\delta)$, as the ${\bf G}$-driven term. Crucially, this interaction arises within the Maxwell sector rather than the fluid sector. Earlier treatments of this phenomenon omitted this channel entirely, operating under the assumption that the interaction was governed solely by ${\bf G}$.
        \item[(ii)] The fluid vorticity acquires a GW-driven piece: Conventionally, the vorticity within the chiral vortical current $\xi\,\bm\omega$ is defined using an ordinary partial derivative, $\omega^\mu=\epsilon^{\mu\nu\alpha\beta}u_\nu\partial_\alpha u_\beta$, rather than a full covariant derivative. This substitution is strictly valid only in a coordinate basis, where the Christoffel symbols $\Gamma^\lambda{}_{\alpha\beta}$ are symmetric in their lower indices. Contracting these symmetric symbols with the totally antisymmetric tensor $\epsilon^{\mu\nu\alpha\beta}$ results in zero, ensuring that $\nabla_\alpha u_\beta$ and $\partial_\alpha u_\beta$ are mathematically equivalent within this specific contraction. However, this cancellation breaks down in a non-coordinate tetrad frame because the spin connection $\omega^c{}_{ab}$ lacks this symmetry. Consequently, the connection-dependent term $\epsilon^{\mu\nu\alpha\beta}u_\nu\,\omega^c{}_{\alpha\beta}u_c$ is non-zero. For a stationary background fluid subjected to transverse velocity perturbations $\delta v_x$ and $\delta v_y$ (where $\delta v_z=0$ and the perturbations depend exclusively on $z-t$), evaluating this connection-driven term results in, at linear order:
       \begin{align}
           \delta\omega^{(1)}{}_{\rm(GW)} &= \tfrac12\big(\dot h_+\,\delta v_y - \dot h_\times\,\delta v_x\big)\, , \\     
           \delta\omega^{(2)}{}_{\rm(GW)} &= \tfrac12\big(\dot h_+\,\delta v_x + \dot h_\times\,\delta v_y\big)\, . 
       \end{align}
        
        This contribution arises in addition to the standard $\bar\nabla\times\delta{\bf v}$ term present in non-relativistic magnetohydrodynamics (MHD). Note that the overall sign of the Levi-Civita tensor, parameterized as $\epsilon^{0123}=s$, is strictly constrained. Requiring the flat-space limit, $\omega^\mu\to\epsilon^{\mu\nu\alpha\beta}u_\nu\partial_\alpha u_\beta$, to reproduce the standard right-handed curl, $\bar\nabla\times\delta{\bf v}$, fixes $s=+1$. This convention aligns with the induction equation, the chiral vortical effect (CVE) term $\xi_0\delta\bm\omega$ utilized throughout this work, and the formulation of Yamamoto's chiral Alfv\'en wave~\cite{yamamoto2015chiral}, remaining completely consistent with the equations above. This connection-induced term emerges at the precise perturbative order, $O(h\,\delta v)$, relevant to the master equations. It is generated entirely by the gravitational wave interacting with a pre-existing velocity perturbation via the anholonomic connection. Because it vanishes when $\delta v_x=\delta v_y=0$, this effect amplifies or rotates existing plasma motion rather than exciting a quiescent plasma from rest. Furthermore, since it couples directly to $\xi\,\bm\omega$. where the transport coefficient $\xi$ is strictly zero for non-chiral plasmas—this channel is uniquely chirality-specific. It possesses no counterpart in ordinary, non-chiral MHD treatments of this configuration and represents a fundamentally novel interaction omitted in earlier studies of this coupling.
    \end{itemize}
    \nocite{*}
	\bibliographystyle{apsrev4-2} 
	\bibliography{ref} 
	
\end{document}